\documentclass[11pt,a4paper]{article}
\pdfoutput=1
\usepackage{amssymb}
\usepackage[final]{graphicx}

\def\blind{0}

\usepackage{newpxtext}
\usepackage[varbb]{newpxmath}
\usepackage{sourcesanspro}

\usepackage{sectsty}
\allsectionsfont{\sffamily}

\usepackage[margin=1in]{geometry}

\usepackage[activate={true,nocompatibility}]{microtype}

\usepackage[font=footnotesize]{caption}

\usepackage{enumitem}

\usepackage{setspace}

\usepackage{hyperref}
\hypersetup{bookmarksopen=true,bookmarksnumbered=true,hidelinks,final=true}

\usepackage{doi}

\usepackage[style=numeric-comp,
            backend=biber,
						natbib=true,
            sorting=none,
            urldate=comp,
            maxcitenames=2,
            mincitenames=1,
            maxbibnames=3,
            giveninits=true]{biblatex}
\usepackage{bm}

\newcommand{\dd}{\mathop{}\!d}

\DeclareMathOperator{\Normal}{N}
\DeclareMathOperator{\halfNormal}{half-N}

\DeclareMathOperator{\Bernoulli}{Bern}
\DeclareMathOperator{\Binomial}{Bin}

\DeclareMathOperator{\Gammadist}{Gam}

\newcommand*{\tr}{^{\scriptscriptstyle\mathsf{T}}}

\usepackage{mleftright}

\usepackage{pdflscape}

\usepackage[capitalise,noabbrev]{cleveref}

\usepackage{subfig}

\usepackage{array,booktabs}

\begin{document}
\singlespacing
\thispagestyle{empty}
\begin{titlepage}
	\begin{center}\LARGE\sffamily\bfseries
		Multilevel network meta-regression for general likelihoods: synthesis of individual and aggregate data with applications to survival analysis
	\end{center}
	\vspace{2\baselineskip}
	\if\blind0
	\begin{raggedright}\footnotesize
		David M.\ Phillippo\footnote{University of Bristol, Canynge Hall, 39 Whatley Road, Bristol, BS8 2PS, UK. Email:	david.phillippo@bristol.ac.uk}\\
		University of Bristol, UK\\[1em]
		Sofia Dias\\
		University of York, UK and University of Bristol, UK\\[1em]
		A.\ E.\ Ades\\
		University of Bristol, UK\\[1em]
		Nicky J.\ Welton\\
		University of Bristol, UK
	\end{raggedright}
	\fi
	\vspace{2\baselineskip}
  \subsection*{Abstract}
	Network meta-analysis combines aggregate data (AgD) from multiple randomised controlled trials, assuming that any effect modifiers are balanced across populations. 
	Individual patient data (IPD) meta-regression is the ``gold standard'' method to relax this assumption, however IPD are frequently only available in a subset of studies. 
	Multilevel network meta-regression (ML-NMR) extends IPD meta-regression to incorporate AgD studies whilst avoiding aggregation bias, but currently requires the aggregate-level likelihood to have a known closed form. 
	Notably, this prevents application to time-to-event outcomes.

	We extend ML-NMR to individual-level likelihoods of any form, by integrating the individual-level likelihood function over the AgD covariate distributions to obtain the respective marginal likelihood contributions. 
	We illustrate with two examples of time-to-event outcomes, showing the performance of ML-NMR in a simulated comparison with little loss of precision from a full IPD analysis, and demonstrating flexible modelling of baseline hazards using cubic M-splines with synthetic data on newly diagnosed multiple myeloma.

	ML-NMR is a general method for synthesising individual and aggregate level data in networks of all sizes. 
	Extension to general likelihoods, including for survival outcomes, greatly increases the applicability of the method. 
	R and Stan code is provided, and the methods are implemented in the \textit{multinma} R package.

	\paragraph{Keywords} network meta-analysis; effect modification; population adjustment; individual patient data; indirect comparison.
\end{titlepage}

\clearpage
\onehalfspacing
\section{Introduction}\label{sec:introduction}
Healthcare decision-making requires reliable estimates of the relative effectiveness of all relevant treatments in a given population.
Standard indirect comparison and network meta-analysis methods are commonly used to synthesise evidence from multiple trials, each of which potentially compares only a subset of the treatments of interest, under the assumption that there is no imbalance in effect-modifying variables between the trials \parencite{Bucher1997,Higgins1996,Lu2004,TSD2}.
However, when effect modification is present these methods may be biased.
The ``gold standard'' approach to adjust for effect modifiers and relax this assumption is network meta-regression with individual patient data (IPD) available for all studies \parencite{Berlin2002,Lambert2002,Riley2010,TSD3}. 
However, this level of data availability is rare---particularly in contexts such as health technology assessment.
Population adjustment methods have therefore been proposed that use IPD from the subset of studies where it is available, and published aggregate data (AgD) from the rest \parencite{TSD18,Phillippo2018}.
A substantial majority of applications of population adjustment analyses to date involve survival or time-to-event data \parencite{Phillippo2019}; however, current population adjustment approaches are faced with significant limitations or have not yet been extended to handle survival data.

Matching-adjusted indirect comparison (MAIC) is a widely-used population adjustment method that re-weights individuals in one IPD study to match the covariate distribution in an AgD study \parencite{Signorovitch2010,Ishak2015,TSD18}.
Since IPD are only available from one of the studies weights are typically estimated using the method of moments (although alternatives have been proposed \parencite{Jackson2020}), which has been shown to be equivalent to an entropy-balancing approach \parencite{Phillippo2020_maiceb}.
Whilst MAIC is currently the most widely-used method for population adjustment with survival data \parencite{Phillippo2019}, it is limited to the pairwise indirect comparison scenario with one IPD study and one AgD study and cannot readily be extended to incorporate larger networks of studies and treatments \parencite{TSD18}.
Moreover, population-adjusted estimates can only be produced for the AgD study population, which may not be representative of the target population for a treatment decision \parencite{TSD18}.

Simulated treatment comparison (STC) is an alternative approach based on regression adjustment, where a regression model fitted in the IPD study is used to predict outcomes on each treatment in the AgD study population \parencite{Caro2010,Ishak2015,TSD18}.
However, when the outcome measure is non-collapsible, such as hazard ratios or odds ratios, the typical ``plug-in means'' approach is biased due to combining incompatible conditional and marginal effect measures (from the IPD and AgD studies, respectively), a form of aggregation bias \parencite{RemiroAzocar2021_comment,Phillippo2020_response_to_RHB}.
Simulation can be used to avoid this bias \parencite{Caro2010}, however this complicates variance estimation.
A more sophisticated form of STC based on G-computation via simulation from the joint covariate distribution in the AgD study has been developed to address this issue, and variance estimation is handled by bootstrapping or embedding in a Bayesian analysis \parencite{RemiroAzocar2022}.
However, like MAIC, all of these approaches are only applicable to pairwise indirect comparisons and cannot produce estimates for target populations other than that represented by the AgD study.

Multilevel network meta-regression (ML-NMR) is a population adjustment method that extends IPD network meta-regression to coherently incorporate evidence from both IPD and AgD sources \parencite{Phillippo2020_methods,Phillippo_thesis}.
Aggregation bias is avoided by integrating the individual-level model over the joint covariate distribution in the AgD studies, in contrast to previous meta-regression approaches \parencite{Sutton2008,Saramago2012,Donegan2013} that combine IPD and AgD by simply ``plugging in'' mean covariate values from the AgD studies.
Unlike MAIC and STC, ML-NMR can coherently synthesise evidence from networks of any size, and crucially for decision-making can produce population-adjusted estimates of relative or absolute effects in any target population of interest.
Moreover, in larger networks key assumptions regarding unobserved effect modifiers and effect modifier interactions can be assessed using ML-NMR, whereas these are untestable assumptions under all approaches when performing pairwise indirect comparisons \parencite{Phillippo2022}.
ML-NMR is an extension of the standard network meta-analysis (NMA) framework \parencite{TSD2,Higgins1996,Lu2004}, reducing to IPD network meta-regression if IPD are available from all studies, and to AgD NMA when no covariates are included in the model.
\textcite{Phillippo2020_methods} construct the aggregate-level model for ML-NMR in two steps: i) deriving the aggregate likelihood from the individual likelihood, using standard results on the sums of random variables; and ii) integrating the individual-level model over the covariate distribution in the aggregate population to form the aggregate-level model, using a general numerical approach based on quasi-Monte Carlo integration.
However, derivation of the aggregate likelihood is not straightforward in general and may even be intractable, since analytic results for the sums of random variables are only available for some special cases (e.g.\ Normal, Poisson, or Bernoulli distributions \parencite{Phillippo2020_methods}, or ordered categorical distributions \parencite{Phillippo2022}).
Most notably this is the case for the analysis of survival outcomes where the aggregate likelihood cannot be derived analytically.

In this paper, we begin by setting out the ML-NMR framework in a more general form based on the likelihood contributions from different sources of data.
We directly integrate the individual-level likelihood function over the joint covariate distribution to obtain the likelihood contributions for the AgD studies, using quasi-Monte Carlo integration.
This approach does not require the form of the aggregate-level likelihood to be analytically tractable, or even known.
We then use this approach to describe ML-NMR models for censored time-to-event outcomes with general survival and hazard functions.
Finally, we apply these ideas to two examples of survival outcomes, one simulated comparison showing performance against full IPD network meta-regression in recovering true parameter values, and another demonstrating flexible modelling of survival with synthetic data on newly diagnosed multiple myeloma.

\section{ML-NMR for general likelihoods}\label{sec:methods}
Consider the general network meta-analysis setting, where we have $J$ randomised controlled trials, each investigating a subset $\mathscr{K}_j$ of $K$ treatments.
If IPD are available from each of the $K$ studies, then we can estimate a standard IPD network meta-regression model, which may be written as
\begin{subequations}\label{eqn:IPD_NMA}
\begin{align}
	y_{ijk} &\sim \pi_{\mathrm{Ind}}(\theta_{ijk}) \\
	g(\theta_{ijk}) &= \eta_{jk}(\bm{x}_{ijk}) = \mu_j + \bm{x}_{ijk}\tr (\bm{\beta}_1 + \bm{\beta}_{2,k}) + \gamma_k
\end{align}
\end{subequations}
with IPD outcomes $y_{ijk}$ for individuals $i=1,\dots,N_{jk}$ in study $j=1,\dots,J$ receiving treatment $k\in\mathscr{K}_j$ given the likelihood distribution $\pi_{\mathrm{Ind}}(\theta_{ijk})$.
The link function $g(\cdot)$ links the likelihood parameter $\theta_{ijk}$ to the linear predictor $\eta_{jk}(\bm{x}_{ijk})$, with covariates $\bm{x}_{ijk}$.
The parameters $\mu_j$ are study-specific intercepts, $\bm{\beta}_1$ and $\bm{\beta}_{2,k}$ are regression coefficients for prognostic and effect modifying covariates respectively, and $\gamma_k$ are individual-level treatment effects.
We set $\bm{\beta}_{2,1}=\gamma_1=0$ for the reference treatment 1.

By specifying an individual-level model \eqref{eqn:IPD_NMA}, with a likelihood, link function, and linear predictor, we are also specifying an individual-level likelihood function, conditional on the covariate values for each individual.
Letting $\bm{\xi}$ denote the set of all model parameters $\{ \mu_j, \bm{\beta}_1, \bm{\beta}_{2,k}, \gamma_k : \forall j, k \}$, we denote the individual conditional likelihood function by $L^\mathrm{Con}_{ijk | \bm{x}}(\bm{\xi} ; y_{ijk}, \bm{x}_{ijk})$.
The form of this individual conditional likelihood function follows from the chosen individual-level model, in particular the individual-level likelihood $\pi_{\mathrm{Ind}}(\cdot)$, link function $g(\cdot)$, and linear predictor $\eta_{jk}(\cdot)$.

To extend the IPD network meta-regression model \eqref{eqn:IPD_NMA} into a ML-NMR model that incorporates aggregate-level evidence from studies where IPD are not available, we integrate the individual conditional likelihood function over the joint covariate distribution in an AgD study to obtain an individual \emph{marginal} likelihood function, describing the likelihood where individual outcomes are known but individual covariates are not (only summary covariate distributions).
For example, this is the case when analysing survival outcomes using time-to-event data reconstructed from published Kaplan-Meier curves but with only published summary covariate information at baseline.
Integrating the individual conditional likelihood function over the joint covariate distribution $f_{jk}(\cdot)$ on treatment $k$ in study $j$, we obtain the individual marginal likelihood function
\begin{equation} \label{eqn:gl_individual_marg_lik}
  L^\mathrm{Mar}_{ijk}(\bm{\xi} ; y_{ijk}) = \int_{\mathfrak{X}} L^\mathrm{Con}_{ijk | \bm{x}}(\bm{\xi} ; y_{ijk}, \bm{x}) f_{jk}(\bm{x}) \dd\bm{x},
\end{equation}
which no longer depends on $\bm{x}$.
In other words, for an individual on treatment $k$ in study $j$ with outcome $y_{ijk}$, if we do not know their individual covariate vector $\bm{x}_{ijk}$ but only the distribution $f_{jk}(\cdot)$, their likelihood contribution is given by \eqref{eqn:gl_individual_marg_lik}.
This integration may be performed using quasi-Monte Carlo integration, as described previously \parencite{Phillippo2020_methods}.
With a set of $\tilde{N}$ integration points $\tilde{\bm{x}}_{jk}$ drawn from $f_{jk}(\cdot)$, the individual marginal likelihood function \eqref{eqn:gl_individual_marg_lik} is evaluated as
\begin{equation}\label{eqn:gl_individual_marg_lik_numint}
  L^\mathrm{Mar}_{ijk}(\bm{\xi} ; y_{ijk}) \approx \tilde{N}^{-1} \sum_{\tilde{\bm{x}}} L^\mathrm{Con}_{ijk | \bm{x}}(\bm{\xi} ; y_{ijk}, \bm{x}).
\end{equation}
In practice, it is likely that only marginal covariate summaries are available from the AgD studies instead of the full joint distribution $f_{jk}(\cdot)$, but we can reconstruct the joint distribution given assumed forms for the marginal covariate distributions and the correlation matrix, for example assuming that these are the same as those observed in the IPD studies \parencite{Phillippo2020_methods}.
Simulation studies with binary outcomes have found that the results of ML-NMR analyses are not sensitive to the assumptions used in reconstructing the joint distribution \parencite{Phillippo2020_simstudy}; we expect this result to hold for other outcomes including survival, and indeed have found this to be the case in our experience.

If we have summary outcomes $y_{\bullet jk}$ on a given treatment $k$ in study $j$, we can attempt to derive a corresponding \emph{aggregate} marginal likelihood function as the product of the individual marginal likelihood functions \eqref{eqn:gl_individual_marg_lik}, up to a normalising constant:
\begin{equation} \label{eqn:gl_aggregate_lik}
  L^\mathrm{Mar}_{\bullet jk}(\bm{\xi} ; y_{\bullet jk}) \propto \prod_{i=1}^{N_{jk}} L^\mathrm{Mar}_{ijk}(\bm{\xi} ; y_{ijk}),
\end{equation}
where the subscript $\bullet$ denotes quantities that have been aggregated over individuals.
If the result can be rearranged in terms of $y_{\bullet jk}$, we can then use $L^\mathrm{Mar}_{\bullet jk}(\bm{\xi} ; y_{\bullet jk})$ to evaluate the aggregate marginal likelihood function.
This is possible when outcomes are discrete (e.g. binary outcomes, as we demonstrate with some discussion in \cref{ssec:gl_binary_derivation}), but may not be possible in general.

By working directly with the likelihood contributions from each level of the model, we avoid having to explicitly derive the form of the aggregate likelihood.
The full ML-NMR model for general likelihoods may be written using \eqref{eqn:gl_individual_marg_lik} and \eqref{eqn:gl_aggregate_lik} as
\begin{subequations}\label{eqn:gl_ML-NMR_model}
\begin{align}
  \intertext{Individual:}
  L^\mathrm{Con}_{ijk | \bm{x}}(\bm{\xi} ; y_{ijk}, \bm{x}_{ijk}) &= \pi_{\mathrm{Ind}}(y_{ijk} | \theta_{ijk}) \label{eqn:gl_ML-NMR_model_IPD_lik} \\
	g(\theta_{ijk}) &= \eta_{jk}(\bm{x}_{ijk}) = \mu_j + \bm{x}_{ijk}\tr (\bm{\beta}_1 + \bm{\beta}_{2,k}) + \gamma_k \label{eqn:gl_ML-NMR_model_IPD} \\
  \intertext{Aggregate:}
  L^\mathrm{Mar}_{ijk}(\bm{\xi} ; y_{ijk}) &= \int_{\mathfrak{X}} L^\mathrm{Con}_{ijk | \bm{x}}(\bm{\xi} ; y_{ijk}, \bm{x}) f_{jk}(\bm{x}) \dd\bm{x} \label{eqn:gl_ML-NMR_model_AgD_ind} \\
  L^\mathrm{Mar}_{\bullet jk}(\bm{\xi} ; y_{\bullet jk}) &\propto \prod_{i=1}^{N_{jk}} L^\mathrm{Mar}_{ijk}(\bm{\xi} ; y_{ijk}) \label{eqn:gl_ML-NMR_model_AgD}
\end{align}
\end{subequations}
where in a Bayesian analysis, prior distributions are placed over each of the parameters $\mu_j$, $\bm{\beta}_1$, $\bm{\beta}_{2,k}$, and $\gamma_k$.

Computationally, we fit these models in Stan by directly coding the log likelihood contributions with a \texttt{target}~\texttt{+=} statement \parencite{stan-reference-manual}.
These models may also be fitted in WinBUGS/OpenBUGS/JAGS by using the ``zeros trick'' to provide the correct (log) likelihood contributions via a Poisson distribution with dummy zero observations \parencite{Lunn2010}.

\subsection{Application to survival analysis}\label{ssec:gl_survival_derivation}
We now apply this general framework to derive ML-NMR models for survival or time-to-event outcomes.
We consider the scenario where every study provides a pair $y_{ijk} = \lbrace t_{ijk}, c_{ijk} \rbrace$ of outcome times $t_{ijk}$ and censoring indicators $c_{ijk}$ for each individual $i$ in study $j$ receiving treatment $k$, where $c_{ijk}=1$ if an individual experiences the event or $c_{ijk}=0$ if they are censored.
For the AgD studies, this data could be obtained by digitizing published Kaplan-Meier curves and reconstructing the event and censoring times using an algorithm such as that described by \textcite{Guyot2012}.
Individual covariate information $\bm{x}_{ijk}$ is available for every individual in the IPD studies, but for the AgD studies only the joint distribution of the covariates at baseline $f_{jk}(\cdot)$ is available (or more likely reconstructed from reported marginal summaries \parencite{Phillippo2020_methods}).

The individual conditional likelihood contributions for each time $t_{ijk}$ in the IPD are given by
\begin{equation}\label{eqn:gl_survival_individual_cond_lik}
  L^\mathrm{Con}_{ijk | \bm{x}}(\bm{\xi}; t_{ijk}, c_{ijk}, \bm{x}_{ijk}) = S_{jk}(t_{ijk}|\bm{x}_{ijk}) h_{jk}(t_{ijk}|\bm{x}_{ijk})^{c_{ijk}},
\end{equation}
where $S_{jk}(t|\bm{x})$ and $h_{jk}(t|\bm{x})$ are the survival and hazard functions conditional on covariates $\bm{x}$.
The forms of the survival and hazard functions depend on the specific parametric model chosen, but the framework described here may be applied in any case, as long as both the survival and hazard functions are specified.
For example, a Weibull proportional hazards model has survival and hazard functions
\begin{align*}
	S_{jk}(t|\bm{x}) &= \exp\mleft( -t^{\nu_j} \exp(\eta_{jk}(\bm{x})) \mright)\\
	h_{jk}(t|\bm{x}) &= \nu_j  t^{\nu_j - 1} \exp(\eta_{jk}(\bm{x}))
\end{align*}
where $\nu_j$ is a study-specific shape parameter.
Notice that we stratify the baseline hazard by study to respect randomisation, i.e.\ for the Weibull model the shape parameters $\nu_j$ are study-specific, akin to the stratification of the study-specific intercepts $\mu_j$ in the linear predictor.
\Cref{sec:survival_hazard_functions} details survival and hazard functions for all survival models currently implemented in the \textit{multinma} R package \parencite{multinma}.
These include Exponential and Weibull proportional hazards models (\cref{sec:ph_models}), and Exponential, Weibull, Gompertz, log-Normal, log-Logistic, Gamma, and generalised Gamma accelerated failure time models (\cref{sec:aft_models}).
The \textit{multinma} package also implements a novel flexible baseline hazards model using M-splines, of which piecewise exponential models are a special case (\cref{ssec:mspline_model,ssec:pexp_model}).

Using equation \eqref{eqn:gl_individual_marg_lik}, the individual marginal likelihood contributions for each event/censoring time in the AgD studies are
\begin{equation}\label{eqn:gl_survival_individual_marg_lik}
\begin{aligned}
  L^\mathrm{Mar}_{ijk}(\bm{\xi} ; t_{ijk}, c_{ijk}) &= \int_{\mathfrak{X}} L^\mathrm{Con}_{ijk | \bm{x}}(\bm{\xi}; t_{ijk}, c_{ijk}, \bm{x}) f_{jk}(\bm{x}) \dd\bm{x} \\
  &= \int_{\mathfrak{X}} S_{jk}(t_{ijk}|\bm{x}) h_{jk}(t_{ijk}|\bm{x})^{c_{ijk}} f_{jk}(\bm{x}) \dd\bm{x}.
\end{aligned}
\end{equation}
We evaluate this integral using quasi-Monte Carlo integration following equation \eqref{eqn:gl_individual_marg_lik_numint} as
\begin{equation}\label{eqn:gl_survival_individual_marg_lik_numint}
  L^\mathrm{Mar}_{ijk}(\bm{\xi} ; t_{ijk}, c_{ijk}) \approx \tilde{N}^{-1} \sum_{\tilde{\bm{x}}} S_{jk}(t_{ijk}|\tilde{\bm{x}}) h_{jk}(t_{ijk}|\tilde{\bm{x}})^{c_{ijk}}.
\end{equation}

\subsection{Model comparison}\label{sec:gl_model_comparison}
Model comparison for network meta-analyses fitted in a Bayesian framework is typically performed using the Deviance Information Criterion (DIC) \parencite{Spiegelhalter2002,TSD2}.
However, the general ML-NMR model \cref{eqn:gl_ML-NMR_model} may not have a closed-form aggregate-level likelihood, which means that the usual $p_D$ complexity penalty cannot easily be evaluated.
Instead, the DIC may be calculated using the $p_V$ penalty proposed by \textcite{BDA3}, or more recently proposed information criteria such as the Watanabe-Akaike Information Criterion (WAIC) or Leave-One-Out Information Criterion (LOOIC) \parencite{Vehtari2016} can be used, all of which are calculated directly from the log likelihood contributions.
We choose to use the LOOIC here, as it (along with WAIC as an asymptotic approximation to LOOIC) has a number of advantages over DIC, including that predictive performance is evaluated over the entire posterior distribution rather than only at a point estimate, and LOOIC works well when the posterior is not approximately Normal \parencite{Vehtari2016}.

\subsection{Assessing integration error}\label{ssec:integration_error}
ML-NMR models are typically implemented using Quasi-Monte Carlo integration via Sobol' sequences to evaluate the integral for the aggregate-level model, which has an expected error rate of $1/\tilde{N}$ \parencite{Phillippo2020_methods}.
\textcite{Phillippo2020_methods} previously suggested assessing the accuracy of the numerical integration by plotting the empirical integration error over the entire posterior distribution for increasing values of $\tilde{N}$.
Whilst this approach may be suitable when the aggregate-level model is of the form \eqref{eqn:gl_ML-NMR_model_AgD} and can be simplified into a single integral per AgD study arm (e.g. for the average event probability in a model with Binomial outcomes), it becomes untenable in practice when the aggregate-level model is of the form \eqref{eqn:gl_ML-NMR_model_AgD_ind} and there is one integral for every individual in each AgD study (e.g. survival outcomes with reconstructed Kaplan-Meier data).
In this case, there may be hundreds or even thousands of such individuals and corresponding integration error plots, and the computational burden of saving and plotting the cumulative integration points quickly becomes unfeasibly heavy in both time and memory.

Instead, we propose the following algorithm to ensure that $\tilde{N}$ is sufficient using the $\widehat{R}$ convergence statistic \parencite{Vehtari2020}, based on the usual practice of fitting $C>1$ chains in parallel (usually $C = 4$):
\begin{enumerate}
	\item Let $n=1$. Select an initial number of integration points $\tilde{N}_1$.
	\item Fit the model, running chains $\mathscr{C}_1$ to $\mathscr{C}_{\lceil C/2 \rceil}$ with $\tilde{N}_n$ integration points, and chains $\mathscr{C}_{\lceil C/2 \rceil+1}$ to $\mathscr{C}_{C}$ with $\lceil \tilde{N}_n/2 \rceil$ integration points, where $\lceil\cdot\rceil$ represents the ceiling operator, rounding up to the nearest integer.
	\item For every parameter,
		\begin{enumerate}[nosep]
			\item Calculate $\widehat{R}_A = \widehat{R}(\mathscr{C}_1, \dots, \mathscr{C}_C)$, the $\widehat{R}$ across all chains combined; 
			\item Calculate $\widehat{R}_W = \max(\widehat{R}(\mathscr{C}_1, \dots, \mathscr{C}_{\lceil C/2 \rceil}), \widehat{R}(\mathscr{C}_{\lceil C/2 \rceil + 1}, \dots, \mathscr{C}_{C}))$, the maximum $\widehat{R}$ across chains sharing the same number of integration points.
		\end{enumerate}
	\item 
		\begin{enumerate}[nosep]
			\item If any $\widehat{R}_W > 1.05$ then the MCMC sampler has not converged; repeat from step 2 with a larger number of iterations.
			\item Else if any $\widehat{R}_A > 1.05$ then the numerical integration has not converged; increase $n$ by 1, let $\tilde{N}_n = 2\tilde{N}_{n-1}$ and go to step 2.
			\item Otherwise $\tilde{N}_n$ is adequate.
		\end{enumerate}
\end{enumerate}
Calculation of the $\widehat{R}$ convergence statistic based on the ratio of within- and between-chains standard deviation was first described by \textcite{Gelman1992}; we use the implementation in the \textit{rstan} R package \parencite{rstan} that incorporates a number of improvements to increase the sensitivity of $\widehat{R}$ to different types of non-convergence \parencite{Vehtari2020}.

Each iteration of this algorithm (i.e. doubling $\tilde{N}$) halves the expected integration error.
Values of $\tilde{N}$ that are powers of 2 are recommended, as these are expected to be particularly efficient for numerical integration schemes based on Sobol' points \parencite{Owen2013}.
The sufficient value of $\tilde{N}$ will vary depending on the model, and we have observed suitable values as low as 16 or as high as 256.
In our experience, a value of $\tilde{N}_1 = 64$ strikes a conservative balance between sufficient accuracy and increased runtime, and should be sufficient for many models to only require a single run.
The \textit{multinma} R package \parencite{multinma} implements the above algorithm (with $\tilde{N}_1 = 64$ by default), and provides user-friendly warnings when the number of integration points is detected to be insufficient.

\subsection{Checking model assumptions}\label{ssec:checking_model_assumptions}
The key assumption underlying all anchored population adjustment approaches is \emph{conditional constancy of relative effects}, which requires that there are no unobserved effect modifiers in imbalance between the included study populations and between these and the target population \parencite{TSD18}.
With ML-NMR, we can assess this assumption using standard techniques from the network meta-analysis literature, checking for residual heterogeneity and inconsistency that may indicate a violation of conditional constancy of relative effects \parencite{Phillippo2020_methods,Phillippo2022}.
Residual heterogeneity can be assessed using a random effects model \parencite{TSD2}, replacing $\gamma_k$ in equation \eqref{eqn:gl_ML-NMR_model} by a study-specific random effect $\delta_{jk} \sim \Normal(\gamma_k, \tau^2)$, where $\tau$ is the between-studies standard deviation.
For studies with more than two arms, a multivariate Normal random effects distribution is required to account for the correlation between relative effects \parencite{TSD2,Phillippo2020_methods}.
Residual inconsistency can be assessed using unrelated mean effects or node-splitting models \parencite{TSD4}.
For example, an unrelated mean effects model replaces $\gamma_k$ in equation \eqref{eqn:gl_ML-NMR_model} by $\gamma_{t_{j1}k}$, where $t_{j1}$ is the treatment in arm 1 of study $j$ and we set $\gamma_{kk} = 0$ for all $k$.
\textcite{Phillippo2022} demonstrate the practical application of these techniques to ML-NMR models, all of which are implemented in the \textit{multinma} R package.

In practical applications of ML-NMR, we often find that the available data are insufficient to estimate independent effect modifier interaction terms $\bm{\beta}_{2,k}$ for each treatment.
Where this is the case, we typically rely on the \emph{shared effect modifier assumption} for a set of treatments $\mathscr{T}$, and define the effect modifier interaction terms to be equal for all treatments within this set, $\bm{\beta}_{2,k} = \bm{\beta}_{2,\mathscr{T}} \forall k \in \mathscr{T}$ \parencite{TSD18,Phillippo2020_methods}.
This assumption is likely to be reasonable when treatments belong to the same class, sharing a mode of action \parencite{TSD18}.
\textcite{Phillippo2022} demonstrate how the shared effect modifier assumption may be relaxed and assessed one covariate at a time, which is less data-intensive than fitting a model with independent interactions for all covariates at once.

When fitting time-to-event models, we should also assess the suitability of the proportional hazards assumption (or the analogous accelerated failure time assumption).
We assess this assumption by letting the baseline hazard vary between the arms of each study.
For parametric models like the Weibull model, this means allowing independent shape parameters $\nu_{jk}$ to vary by treatment arm as well as by study.
For a flexible M-spline hazard model (and piecewise constant hazards as a special case), this means allowing independent spline coefficient vectors $\bm{\alpha}_{jk}$ by arm as well as by study.

\subsection{Producing population-average estimates for a target population}\label{ssec:target_poplations}
For decision-making, we must produce estimates of quantities of interest, such as population-average treatment effects or survival probabilities, in a target population relevant to the decision.
The decision target population need not be represented by one of the studies in the network; indeed, it is likely best represented by a registry or cohort study in the population of interest \parencite{TSD18}.

Population-average relative treatment effects $d_{ab(P)}$ between each pair of treatments $a$ and $b$ in a population $P$ can be produced by integrating contrasts of the linear predictor over the joint covariate distribution $f_{(P)}(\bm{x})$, which due to linearity reduces to simply plugging-in mean covariate values $\bar{\bm{x}}_{(P)}$:
\begin{equation}\label{eqn:pop_avg_releff}
\begin{aligned}
	d_{ab(P)} &= \int_{\mathfrak{X}} \mleft( \eta_{(P)b}(\bm{x}) - \eta_{(P)a}(\bm{x}) \mright) f_{(P)}(\bm{x}) \dd\bm{x} \\
		&= \gamma_b - \gamma_a + \bar{\bm{x}}_{(P)}\tr (\bm{\beta}_{2,b} - \bm{\beta}_{2,a})
\end{aligned}
\end{equation}

The primary marginal quantity of interest is the estimated population-average marginal survival function, also called the standardised survival function, from which we can also produce a range of other marginal estimates.
The population-average marginal survival probability $\bar{S}_{(P)k}(t)$ on treatment $k$ in population $P$ at time $t$ is found by integrating the individual-level survival function $S_{(P)k}(t | \bm{x})$ over the joint covariate distribution $f_{(P)}(\bm{x})$ at each time $t$:
\begin{equation}\label{eqn:pop_avg_surv}
	\bar{S}_{(P)k}(t) = \int_{\mathfrak{X}} S_{(P)k}(t | \bm{x}) f_{(P)}(\bm{x}) \dd\bm{x}
\end{equation}
This integral over the joint covariate distribution in the target population can be calculated using the same quasi-Monte Carlo numerical integration approach described earlier, using a set of integration points drawn from the joint distribution $f_{(P)}(\bm{x})$, analogously to \eqref{eqn:gl_individual_marg_lik_numint}.
In the likely scenario that only marginal covariate summaries are available, again we can reconstruct the joint covariate distribution from assumed forms for the marginal distributions and correlation matrix \parencite{Phillippo2020_methods}.
We also require information on the distribution of the baseline hazard in the target population $P$, that is distributions for the linear predictor intercept parameter $\mu_{(P)}$ and any additional parameters of the survival function such as the Weibull shape parameter $\nu_{(P)}$ or M-spline coefficients $\bm{\alpha}_{(P)}$.
Estimates of these parameters may not be available directly for an external target population.
If instead we have (reconstructed) Kaplan-Meier data available for outcomes on a reference treatment in the target population (along with the summary covariate distribution), then this data may be included in the model as a single-arm study at the synthesis stage through equation \eqref{eqn:gl_survival_individual_marg_lik}; this will allow the parameters of the baseline hazard in this population to be estimated, but will not contribute information to any other model parameters.
Otherwise, estimates may be borrowed from a study in the network where the properties of the baseline hazard are deemed to be representative of the target population.

From this marginal survival function, we can then produce a range of other marginal estimates.
The population-average marginal hazard function corresponding to this population-average marginal survival function is a weighted average of the individual-level hazard functions
\begin{equation}\label{eqn:pop_avg_haz}
	\bar{h}_{(P)k}(t) = \frac{\int_\mathfrak{X} S_{(P)k}(t | \bm{x}) h_{(P)k}(t | \bm{x}) f_{(P)k}(\bm{x}) \dd\bm{x} }{\bar{S}_{(P)k}(t)}
\end{equation}
weighted by the probability of surviving to time $t$.
Again, this integral can be calculated using quasi-Monte Carlo numerical integration.
The corresponding population-average marginal cumulative hazard function is
\begin{equation}
	\bar{H}_{(P)k}(t) = -\log\mleft(\bar{S}_{(P)k}(t)\mright).
\end{equation}

Quantiles and medians of the population-average marginal survival times are found by solving
\begin{equation}\label{eqn:pop_avg_quantiles}
	\bar{S}_{(P)k}\mleft(t^{(\alpha)}_{(P)k}\mright) = 1-\alpha	
\end{equation}
to find $t^{(\alpha)}_{(P)k}$ for the $\alpha$\% quantile, which can be achieved using numerical root finding.

Means or restricted means of the population-average marginal survival times are found by integrating the marginal survival function up to a restricted time horizon $t^*$
\begin{equation}\label{eqn:pop_avg_rmst}
	\mathrm{RMST}_{(P)k}(t^*) = \int_0^{t^*} \bar{S}_{(P)k}(t) \dd t
\end{equation}
with $t^*=\infty$ for population-average mean marginal survival time, which is typically evaluated using quadrature; we use the implementation in the \textit{flexsurv} R package \parencite{flexsurv}.

Contrasts of the above quantities may also be created, to form estimates of population-average marginal treatment effects $\Delta_{ab(P)}(t)$.
For example, the population-average marginal hazard functions in equation \eqref{eqn:pop_avg_haz} for two treatments $a$ and $b$ can be combined to form a population-average marginal hazard ratio:
\begin{equation}\label{eqn:pop_avg_marginal_hr}
	\Delta^\mathrm{HR}_{ab(P)}(t) = \frac{\bar{h}_{(P)b}(t)}{\bar{h}_{(P)a}(t)}.
\end{equation}
In a similar fashion we can also create population-average median survival time ratios or differences, or differences in population-average (restricted) mean survival times.

All of the quantities \eqref{eqn:pop_avg_surv} to \eqref{eqn:pop_avg_marginal_hr} are \emph{marginal}, as these are all derived from the population-average marginal survival function $\bar{S}_{(P)k}(t)$.
These quantities all depend on the distributions of the baseline hazard and of all covariates (not just those that are effect-modifying).
Furthermore, we note in particular that the population-average marginal hazard ratios $\Delta^\mathrm{HR}_{ab(P)}(t)$ also vary over time; the presence of covariates (either prognostic or effect modifying) means that, mathematically, proportional hazards \emph{cannot} hold at the marginal level.
In contrast, the $d_{ab(P)}$ are population-average \emph{conditional} treatment effects which depend only on the distribution of \emph{effect-modifying} covariates in the target population.
The $d_{ab(P)}$ are constant over time, and do not depend on the distribution of baseline hazard or the distribution of purely prognostic covariates.
The population-average conditional treatment effects can be interpreted as the average effect between randomly-selected individuals on treatments $a$ and $b$ in the target population $P$ with the same covariates; the population-average marginal treatment effects can be interpreted as the average effect between randomly-selected individuals regardless of their covariates \parencite{Kahan2014}.

\section{Simulated example}\label{sec:gl_example_survival}
To illustrate the performance of this approach, let us consider an artificial example of simulated survival outcomes in a population-adjusted indirect comparison of two treatments $B$ and $C$ via a common comparator $A$.
Since the data are simulated, we can compare the results and performance of ML-NMR using only partial IPD to that of a full IPD NMA, and to the known true values.
We simulate outcomes from a Weibull model including three covariates (two continuous and one binary); full details are given in \cref{sec:simulated_example_details}.

\subsection{Simulated example: Methods}
We fit Exponential, Weibull, and Gompertz proportional hazards models (\cref{sec:ph_models}) in the general ML-NMR framework, each with the linear predictor \eqref{eqn:sim_linear_predictor}, and use the LOOIC to select the most appropriate model.
For comparison, we also fit the corresponding IPD NMA models with full IPD (i.e.\ individual outcomes and covariates) available from both studies.
We also perform a standard (non-population adjusted) indirect comparison, formed from the log hazard ratios estimated in each study separately using a Weibull model without adjustment for effect modifiers but with adjustment for prognostic factors, reflecting ``best case'' common practice (i.e.\ correct form of parametric model, fully adjusted for prognostic factors).

We fit all models in a Bayesian framework, with non-informative $\Normal(0, 100^2)$ prior distributions on every parameter in the linear predictor, and a weakly-informative $\halfNormal(0, 10^2)$ prior distribution on the shape parameter for Weibull and Gompertz models.

Analyses were carried out in R version 4.3.1 \parencite{Rcore} and Stan version 2.26.23 \parencite{Carpenter2017}.
Two sets of analysis codes are provided in the supplementary material that both achieve the results presented here: one that fits the models via the user-friendly \textit{multinma} R package \parencite{multinma}, making these techniques accessible to a broad audience; and another that fits the models by calling Stan directly, which is likely to be useful for those who wish to modify or extend the code for their own purposes.
Using \textit{multinma}, the ML-NMR models take around 3 minutes each to fit on a modern laptop; the IPD NMA models take around 8 seconds each.

\subsection{Simulated example: Results}
Inspecting the LOOIC model comparison statistics in \cref{tab:looic_comparison}, we see that the Weibull model has the lowest LOOIC for both ML-NMR and IPD NMA, and the standard error of the difference suggests that the Weibull model is a substantially better fit than either the Exponential or Gompertz models in both the ML-NMR and IPD NMA scenarios.
Comparing individual LOOIC contributions between the ML-NMR and IPD NMA models reveals that individual observations are fitted similarly well under each model (\cref{fig:survival_looic_contributions}).

\begin{figure}[tbp]
	\centering
	\includegraphics[width=\textwidth]{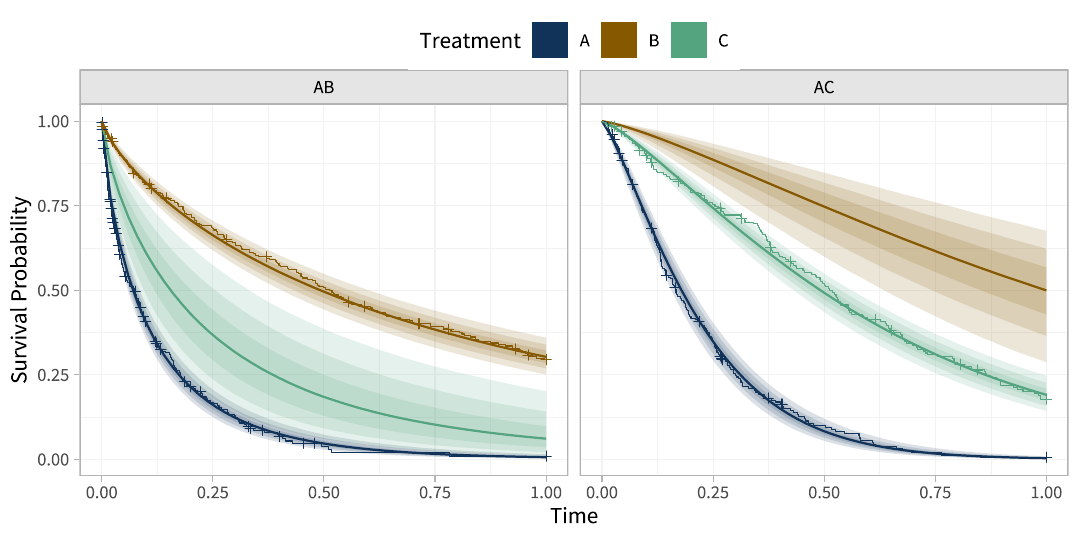}
	\caption{ML-NMR estimated survival curves on each treatment in each study population, under a Weibull model. Shaded bands indicate the 50\%, 80\%, and 95\% Credible Intervals for the survival curves (thick lines), overlaid on the unadjusted Kaplan-Meier curves from the treatments in each study (thin lines).}
	\label{fig:survival_est_curves}
\end{figure}

The estimated population-average survival curves on each treatment in each study population under the Weibull model fitted using ML-NMR are shown in \cref{fig:survival_est_curves}, overlaid on the unadjusted Kaplan-Meier curves.
Visually, the estimated survival curves are a good fit to the observed data.
\Cref{tab:survival_logHR} presents the estimated population-average conditional log hazard ratios (HRs) for each pairwise comparison in each population, along with the true values from the simulation.
The ML-NMR estimates agree well with both the IPD NMA and the true values, and the $B$ vs.\ $A$ and $C$ vs.\ $A$ estimates within the $AB$ and $AC$ study populations respectively are unchanged in point estimate or standard error.
Standard errors for comparisons not observed in the data are slightly increased (by 2--6\%) using ML-NMR compared to full IPD NMA, which is expected due to the reduced information available.

Due to non-collapsibility, we cannot directly compare the estimated log hazard ratios between the population-adjusted models (ML-NMR and IPD NMA) and the unadjusted standard indirect comparison.
Instead, we choose to compare the restricted mean survival times up until the end of follow up ($t^*=1$) on each treatment in each study population under each method, which are displayed in \cref{tab:survival_rmst}.
Since the restricted mean survival time has the same interpretation as a marginal quantity under each of the three models this is a valid comparison.
The results from the ML-NMR and IPD NMA agree closely, with nearly identical posterior means and credible intervals; the estimates of treatment $B$ in the $AC$ population and treatment $C$ in the $AB$ population are slightly more uncertain from the ML-NMR model due to the reduced information available.
However, the standard indirect comparison produces estimates that are clearly biased in this scenario: differences in effect modifiers between the populations are not accounted for, and as a result the difference in restricted mean survival time between treatments $B$ and $C$ is underestimated in both populations.

\begin{table}[tbp]
	\footnotesize
	\centering
	\caption{Table of estimated population-average conditional log hazard ratios and 95\% Credible Intervals from the ML-NMR model and the full IPD NMA, alongside the true log hazard ratios, in the $AB$ and $AC$ study populations.}
	\label{tab:survival_logHR}
	\begin{tabular}{llccc}
		\toprule
		& & \multicolumn{3}{c}{Comparison} \\
		\cmidrule{3-5}
		Study & Method & $B$ vs. $A$ & $C$ vs. $A$ & $C$ vs. $B$ \\
		\midrule
		 AB & Truth & $-1.62$ & $-0.92$ & $0.70$ \\ 
   [0.5ex] & ML-NMR & $-1.53$ & $-0.62$ & $0.90$ \\ 
   &  & ($-1.74$, $-1.30$) & ($-1.19$, $-0.06$) & ($0.28$, $1.52$) \\ 
   [0.5ex] & IPD NMA & $-1.54$ & $-0.67$ & $0.87$ \\ 
   &  & ($-1.76$, $-1.32$) & ($-1.12$, $-0.23$) & ($0.36$, $1.37$) \\ 
   [0.5ex]AC & Truth & $-2.07$ & $-1.37$ & $0.70$ \\ 
   [0.5ex] & ML-NMR & $-2.20$ & $-1.29$ & $0.90$ \\ 
   &  & ($-2.76$, $-1.63$) & ($-1.54$, $-1.05$) & ($0.28$, $1.52$) \\ 
   [0.5ex] & IPD NMA & $-2.17$ & $-1.31$ & $0.87$ \\ 
   &  & ($-2.63$, $-1.70$) & ($-1.54$, $-1.08$) & ($0.36$, $1.37$) \\ 
  
		\bottomrule
	\end{tabular}
\end{table}

\begin{table}[tbp]
	\footnotesize
	\centering
	\caption{Table of estimated restricted mean survival times and 95\% Credible Intervals on each treatment from the ML-NMR model, the full IPD NMA, and the standard indirect comparison, in the $AB$ and $AC$ study populations.}
	\label{tab:survival_rmst}
	\begin{tabular}{llccc}
		\toprule
		& & \multicolumn{3}{c}{Treatment} \\
		\cmidrule{3-5}
		Study & Method & $A$ & $B$ & $C$ \\
		\midrule
		 AB & ML-NMR & $0.13$ & $0.54$ & $0.27$ \\ 
   &  & ($0.11$, $0.16$) & ($0.49$, $0.58$) & ($0.15$, $0.44$) \\ 
   [0.5ex] & IPD NMA & $0.13$ & $0.54$ & $0.28$ \\ 
   &  & ($0.11$, $0.16$) & ($0.49$, $0.58$) & ($0.17$, $0.42$) \\ 
   [0.5ex] & Standard IC & $0.13$ & $0.54$ & $0.47$ \\ 
   &  & ($0.11$, $0.16$) & ($0.49$, $0.58$) & ($0.38$, $0.56$) \\ 
   [0.5ex]AC & ML-NMR & $0.22$ & $0.75$ & $0.53$ \\ 
   &  & ($0.20$, $0.25$) & ($0.62$, $0.84$) & ($0.49$, $0.58$) \\ 
   [0.5ex] & IPD NMA & $0.22$ & $0.74$ & $0.53$ \\ 
   &  & ($0.20$, $0.25$) & ($0.63$, $0.82$) & ($0.49$, $0.57$) \\ 
   [0.5ex] & Standard IC & $0.23$ & $0.59$ & $0.54$ \\ 
   &  & ($0.20$, $0.25$) & ($0.52$, $0.66$) & ($0.50$, $0.58$) \\ 
  
		\bottomrule
	\end{tabular}
\end{table}

Examining the parameters from the ML-NMR and IPD NMA models in \cref{tab:survival_parameter_ests}, we see that these agree closely with each other and recover the true parameter values well.

\section{Newly diagnosed multiple myeloma}\label{sec:example_ndmm}
\begin{figure}
	\centering
  \includegraphics[width=0.8\textwidth]{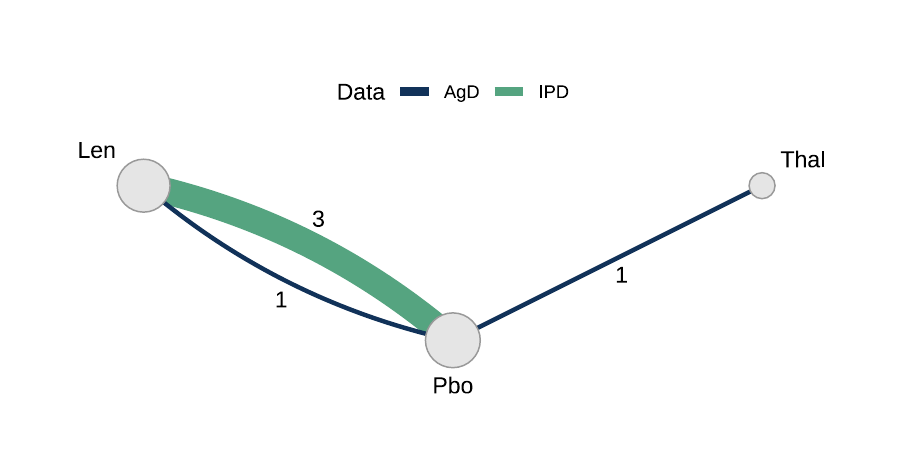}
  \caption{Network of five studies comparing lenalidomide or thalidomide to placebo for treatment of newly diagnosed multiple myeloma. IPD were available from three studies, and AgD from two studies. Edge widths and numbers indicate the number of studies making each comparison, and the size of each node corresponds to the number of individuals randomised to each treatment.}
  \label{fig:ndmm_network}
\end{figure}

To apply these methods to a real example, we consider the analysis of a network of five studies comparing lenalidomide to placebo or thalidomide to  placebo as maintenance treatment for newly diagnosed multiple myeloma \parencite{Leahy2019}.
The outcome of interest is progression-free survival after autologous stem cell transplant (ASCT).
As shown in \cref{fig:ndmm_network}, IPD as individual event/censoring times and covariates are available from three studies; AgD as event/censoring times from digitised Kaplan-Meier curves and overall covariate summaries are available from two studies.
This network was previously analysed by \textcite{Leahy2019}, who applied multiple MAIC analyses before combining in a NMA.
However, there are several disadvantages with this approach, each of which are addressed by ML-NMR; in particular, only the IPD studies are adjusted and the constancy of relative effects assumption is still required when combining the AgD studies, and estimates can only be produced for some weighted-average of the AgD study populations.
We comment on these issues further in the discussion.

Since we did not have access to the original IPD from the three IPD studies, for illustration we instead constructed synthetic data that resemble the original IPD using published Kaplan-Meier curves and regression coefficients.
This process is detailed in \cref{ssec:synthetic_data}.

\subsection{Newly diagnosed multiple myeloma: Methods}
Instead of making parametric assumptions about the form of the baseline hazard, we propose a novel approach using M-splines to flexibly model the baseline hazard over time.
This approach builds on previous applications of M-splines for flexible baseline hazard models in contexts outside of network meta-analysis and ML-NMR \parencite{Brilleman2020,Jackson_survextrap}, and is described in detail in \cref{ssec:mspline_model}.

The survival and hazard functions for the M-spline model are given by
\begin{subequations}
\begin{align}
	S_{jk}(t|\bm{x}) &= \exp\mleft( -\bm{\alpha}_j\tr \bm{I}(t, \bm{\zeta}_j) \exp(\eta_{jk}(\bm{x})) \mright)\\
	h_{jk}(t|\bm{x}) &= \bm{\alpha}_j\tr \bm{M}(t, \bm{\zeta}_j) \exp(\eta_{jk}(\bm{x}))
\end{align}
\end{subequations}
where $\bm{\alpha}_j$ is a study-specific vector of spline coefficients, $\bm{M}_\kappa(t, \bm{\zeta}_j)$ is the M-spline basis of order $\kappa$ with a study-specific knot sequence $\bm{\zeta}_j$ evaluated at time $t$, and $\bm{I}_\kappa(t, \bm{\zeta}_j)$ is the corresponding integrated M-spline basis (an I-spline basis; see \cref{ssec:mspline_model}).
The basis polynomials have degree $\kappa-1$, so a basis of order $\kappa=4$ corresponds to a cubic M-spline basis; a piecewise exponential baseline hazards model is a special case with degree zero ($\kappa=1$).

To avoid overfitting, we propose a novel weighted random walk prior distribution on the inverse-softmax transformed spline coefficients:
\begin{subequations}
\begin{align}
	\bm{\alpha}_j &= \operatorname{softmax}(\bm{\alpha_j^*}) \\
	\alpha^*_{j,l} &= c_{j,l} + \sum_{m = 1}^l u_{j,m} \quad \forall l=1,\dots,L+\kappa-1\\
	u_{j,l} &\sim \Normal(0, \sigma^2_j w_{j,l}) \quad \forall l=1,\dots,L+\kappa-1
\end{align}
\end{subequations}
where $L$ is the number of internal knots.
The random walk is centred around a prior mean vector $\bm{c}_j$ that corresponds to a constant baseline hazard (see \cref{ssec:mspline_model}), borrowing an idea of \textcite{Jackson_survextrap} who derived $\bm{c}_j$ to use instead for the prior mean of a random effect on $\bm{\alpha}_j$.
The weights $w_{j,l}$ are derived from the distance between each pair of knots (see \cref{ssec:mspline_model}), following a similar approach to the Bayesian P-splines proposed by \textcite{Li2022} except that we additionally normalise the weights to sum to 1.
The weights serve to make the prior invariant to the number and location of the knots, even if they are unevenly spaced, and to the timescale, greatly simplifying the specification of a hyperprior distribution for the random walk standard deviation $\sigma_j$.
The random walk standard deviation $\sigma_j$ controls the amount of smoothing and shrinkage of the spline coefficients; as $\sigma_j$ approaches zero the baseline hazard becomes smoother (less ``wiggly'') and approaches a constant baseline hazard.
We allow $\sigma_j$ to be estimated from the data, giving this a weakly-informative hyperprior distribution $\sigma_j \sim \halfNormal(0, 1^2)$.

We adjust for four clinically-relevant covariates considered to be potential effect modifiers by \textcite{Leahy2019}: age, international staging system (ISS) stage (stage III vs.\ stage I-II), response post-ASCT (complete response or very good partial response vs.\ other), and sex (male or female).
The distributions of these covariates in each study at baseline are given in \cref{tab:ndmm_baseline_covariates}.
Due to the lack of data on thalidomide (only a single AgD study), we make the shared effect modifier assumption between the two active treatments in order to identify the effect modifying treatment-covariate interactions \parencite{TSD18,Phillippo2020_methods}.
Since thalidomide and lenalidomide are in the same class of treatments, this assumption may be reasonable.

We fit a cubic M-spline model with seven internal knots placed at evenly-spaced quantiles of the uncensored survival times in each study, plus boundary knots at time 0 and the last event/censoring time in each study.
The number of knots is set to be larger than we might expect to need, since any potential for overfitting is avoided by shrinkage through the random walk prior.
To ensure that seven knots are sufficient, we also fit a model with ten internal knots for comparison.
We assess the proportional hazards assumption by fitting models with spline coefficients $\bm{\alpha}_{jk}$ stratified by treatment arm as well as by study.
We give non-informative $\Normal(0, 100^2)$ prior distributions to every parameter in the linear predictor.
We also fit unadjusted NMA models with the same M-spline baseline hazard for comparison.

Analyses were carried out in R version 4.3.1 \parencite{Rcore} and Stan version 2.26.23 \parencite{Carpenter2017}.
Once again, two sets of analysis codes are provided in the supplementary material that both achieve the results presented here: one that fits the models via \textit{multinma} R package \parencite{multinma}, and another that fits the models by calling Stan directly.
Additionally, the synthetic data are available in the \textit{multinma} R package along with a vignette that walks through the analysis \parencite{multinma}.
Using \textit{multinma}, the ML-NMR models take around 2 hours each to fit on a modern laptop; the unadjusted NMA models take around 7 minutes each.

\subsection{Newly diagnosed multiple myeloma: Results}

\begin{figure}[tbp]
	\centering
	\includegraphics[width=\textwidth]{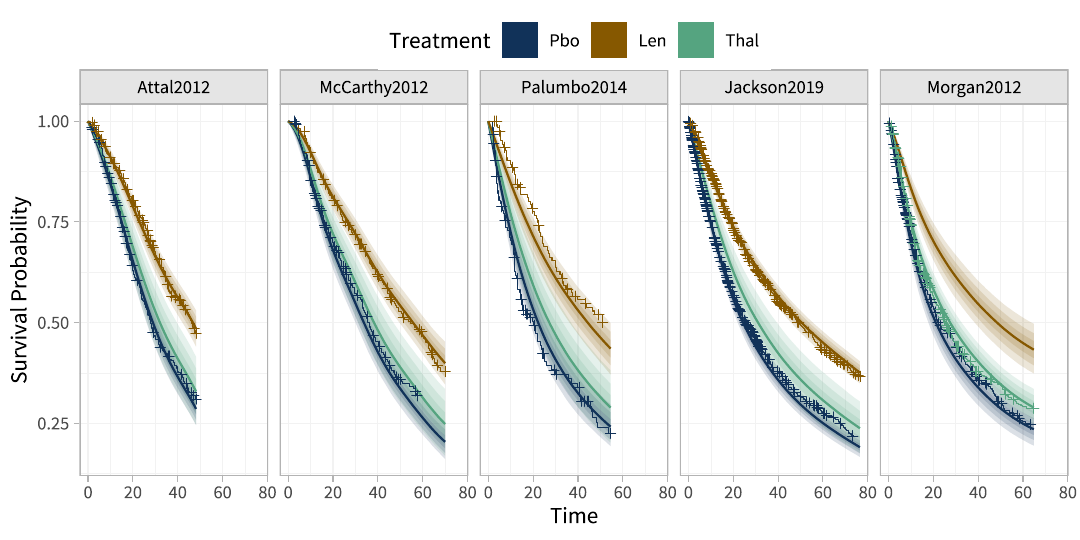}
	\caption{Estimated survival curves on each treatment in each study population, under a cubic M-spline model. Shaded bands indicate the 50\%, 80\%, and 95\% Credible Intervals for the survival curves (thick lines), overlaid on the unadjusted Kaplan-Meier curves from the treatments in each study (thin lines).}
	\label{fig:ndmm_survival_est_curves}
\end{figure}

The estimated population-average survival curves in each study population are shown in \cref{fig:ndmm_survival_est_curves}, overlaid with the observed (unadjusted) Kaplan-Meier curves.
These show a good visual fit to the observed data, with the possible exception of the lenalidomide arm of Palumbo~2014 where the unadjusted Kaplan-Meier estimate lies consistently above the population-adjusted estimate.
This is likely due to the slight baseline imbalance in Palumbo~2014 between arms, with the lenalidomide arm having 17\% fewer males than the placebo arm.
The unadjusted Kaplan-Meier curves do not account for this difference, whereas the population-adjusted survival estimates from the ML-NMR model do.
The population-average median survival times corresponding to these population-average survival curves are given in \cref{tab:ndmm_median_survival}.
The posterior means for the median survival estimates vary across populations between 20.75 and 33.30 months on placebo, 26.55 and 38.44 months on thalidomide, and 44.95 and 55.92 months on lenalidomide.

To assess whether seven internal knots are sufficient, we also fit a model with ten internal knots.
Comparing the model fit in \cref{tab:ndmm_looic_comparison}, we find that there is no substantial difference between the models.
The LOOIC is slightly worse for the model with ten knots, but not substantially, due to a slight increase in the effective number of parameters $p_\mathrm{LOO}$; however, the random walk prior distribution is behaving as expected and controlling the overall complexity through shrinkage. 
This is also apparent when looking at the individual-level baseline hazard functions (\cref{fig:ndmm_baseline_hazard_est_curves}) and the corresponding population-average marginal hazard functions (\cref{fig:ndmm_hazard_est_curves}) which are very similar between models.
We also check the LOOIC within each study separately (\cref{tab:ndmm_looic_by_study}) to ensure that no studies individually are better fit with a higher number of knots, which could be missed when looking overall.
We conclude that seven internal knots are sufficient, both overall and within each study in the network.

To assess the proportional hazards assumption, we modify the M-spline model to stratify the spline coefficients $\bm{\alpha}_{jk}$ on the baseline hazard by treatment arm as well as by study.
Comparing the overall model fit between the models with and without the proportional hazards assumption (\cref{tab:ndmm_looic_nph}), we see that the LOOIC is lower for the proportional hazards model. 
Again, we also check the LOOIC within each study separately (\cref{tab:ndmm_looic_nph_by_study}), to ensure that the proportional hazards assumption is reasonable within each study in the network.
The LOOIC is lower or not substantially higher for the proportional hazards model within each study.
We conclude that the proportional hazards assumption is reasonable here.
For comparison, we also fitted unadjusted models with no covariates (i.e. a standard network meta-analysis) both with and without the proportional hazards assumption.
Whilst there was little difference in the overall model fit (\cref{tab:ndmm_looic_nma}), the non-proportional hazards model did have a substantially lower LOOIC in the Jackson~2019 study (\cref{tab:ndmm_looic_nma_by_study}).
Including the covariates in the ML-NMR analysis, even though they are fixed and not time-varying, is sufficient to remove this proportional hazards violation, and the ML-NMR model is a much better overall fit than the unadjusted NMA.

The estimated population-average conditional log hazard ratios from the ML-NMR model (with seven internal knots and proportional hazards) are given in \cref{tab:ndmm_logHR}.
Both lenalidomide and thalidomide are consistently estimated to be more effective than placebo in each of the study populations, however the 95\% credible intervals for the thalidomide vs.\ placebo comparison crosses zero in both AgD study populations (Jackson~2019 and Morgan~2012), where both relative effects vs.\ placebo are estimated with slightly more uncertainty.
The thalidomide vs.\ lenalidomide relative effect estimates are constant across all populations (0.47, 95\% CrI 0.24 to 0.71), due to the shared effect modifier assumption.

\begin{table}[tbp]
	\footnotesize
	\centering
	\caption{Estimated population-average conditional log hazard ratios and 95\% Credible Intervals in each study population from the cubic M-spline model.}
	\label{tab:ndmm_logHR}
	\begin{tabular}{lccc}
		\toprule
		Study & Lenalidomide vs. Placebo & Thalidomide vs. Placebo & Thalidomide vs. Lenalidomide  \\
		\midrule
		 Attal2012 & $-0.59$ & $-0.13$ & $0.47$ \\ 
   & ($-0.74$, $-0.45$) & ($-0.38$, $0.13$) & ($0.23$, $0.69$) \\ 
   [0.5ex]McCarthy2012 & $-0.62$ & $-0.16$ & $0.47$ \\ 
   & ($-0.74$, $-0.51$) & ($-0.38$, $0.07$) & ($0.23$, $0.69$) \\ 
   [0.5ex]Palumbo2014 & $-0.64$ & $-0.18$ & $0.47$ \\ 
   & ($-0.80$, $-0.48$) & ($-0.44$, $0.09$) & ($0.23$, $0.69$) \\ 
   [0.5ex]Jackson2019 & $-0.69$ & $-0.22$ & $0.47$ \\ 
   & ($-0.81$, $-0.57$) & ($-0.43$, $-0.01$) & ($0.23$, $0.69$) \\ 
   [0.5ex]Morgan2012 & $-0.68$ & $-0.22$ & $0.47$ \\ 
   & ($-0.81$, $-0.56$) & ($-0.42$, $-0.01$) & ($0.23$, $0.69$) \\ 
   [0.5ex]
		\bottomrule
	\end{tabular}
\end{table}

\section{Discussion}\label{sec:gl_discussion}

In this paper, we extended the ML-NMR framework to handle general likelihoods where the aggregate-level likelihood may not have a closed form.
This greatly expands the range of models which can be fitted, including time-to-event outcomes which are common in technology appraisals \parencite{Phillippo2019}.
As in \textcite{Phillippo2020_methods}, we began with a fully-specified individual-level model, and considered how to marginalise this model to apply to aggregate data.
However, instead of explicitly deriving the form of the aggregate likelihood via standard results on the sums of random variables, we proceeded by directly integrating the individual conditional likelihood function over the covariate distribution to obtain the individual marginal likelihood function.
This is then used in one of two ways, depending on the data available, with different levels of generality.

Firstly, in settings where the aggregate data consist of individual outcomes but only summary covariate information (such as survival data reconstructed from Kaplan-Meier curves), the aggregate part of the model is fitted directly using the individual marginal likelihood contributions.
In this case, the method is fully general: individual conditional likelihood functions of any form can be integrated numerically to evaluate the individual marginal likelihood function.
We use the efficient and general quasi-Monte Carlo integration approach proposed by \textcite{Phillippo2020_methods}.

Secondly, we have settings where the aggregate data consist of summary outcomes and summary covariate information.
In this case, the individual marginal likelihood contributions are multiplied together to obtain the aggregate marginal likelihood contributions for the summary outcomes.
Evaluation of the aggregate marginal likelihood contributions requires that these can be expressed in terms of the summary outcomes (as demonstrated in \cref{ssec:gl_binary_derivation}), which is only straightforward for discrete outcomes.
This would appear to limit the generality of the approach for continuous outcomes; however, the aggregate-level likelihood has a known closed form for many continuous individual-level likelihoods common in practice \parencite{Phillippo2020_methods}.

When comparing ML-NMR and full IPD NMA in the simulated example, we found close agreement between the results of ML-NMR (using only aggregate covariate information from the $AC$ study) and full IPD NMA, which both successfully recovered the known true values.
Furthermore, the lack of IPD in the $AC$ study did not greatly reduce precision for ML-NMR compared to IPD NMA; the standard errors of population-average log hazard ratios were the same for comparisons observed within each study population, and only slightly increased for the comparisons not observed.
The conclusions of the model selection process were identical for ML-NMR and IPD NMA, in both cases correctly identifying the Weibull model as the most appropriate model from those fitted.
Nevertheless, this scenario is only a single instance.
A full simulation study could further validate the performance of ML-NMR for survival analysis, and investigate the impact of invalid assumptions.
However, we expect the results and conclusions of previous simulation studies on binary outcomes to apply broadly to ML-NMR models of general forms---including for survival analysis \parencite{Phillippo2020_simstudy}.

Whilst ML-NMR and IPD NMA were seen to perform very similarly in the simulated example, the additional IPD available to IPD NMA does offer additional possibilities for analysis.
For example, we required the shared effect modifier assumption in this scenario to identify the ML-NMR model.
In the interests of a fair comparison between ML-NMR and IPD NMA, both methods made use of this assumption in this analysis---which was known to hold due to the simulated setup.
However, IPD NMA could relax this assumption and estimate separate effect modifier interaction coefficients $\bm{\beta}_{2,B}$ and $\bm{\beta}_{2,C}$, rather than assuming equality.
In this scenario, since we know that $\bm{\beta}_{2,B} = \bm{\beta}_{2,C}$, the standard errors for IPD NMA would have been inflated by the unnecessarily more flexible model.
The shared effect modifier assumption was also used in the newly diagnosed multiple myeloma example, again due to insufficient data to estimate separate treatment-covariate interactions (for thalidomide).
In this case the assumption may be reasonable, since lenalidomide and thalidomide both belong to the same class of treatments.
However, when treatments are not in the same class this assumption is likely to be much less plausible \parencite{TSD18}.
Even when this assumption does not hold, we still expect population-average estimates in the AgD study population to be unbiased \parencite{Phillippo2020_simstudy}.
In larger treatment networks it can be possible to assess and relax the shared effect modifier assumption in ML-NMR \parencite{Phillippo2022}.
When studies across the network report relative effect estimates within subgroups, network meta-interpolation has recently been proposed to combine these in a manner that relaxes the shared effect modifier assumption \parencite{Harari2023}.
Ongoing work aims to utilise subgroup results and regression estimates, where available from trial reports, to support the estimation of ML-NMR models and reduce reliance on the shared effect modifier assumption in practical applications.

When working with a non-collapsible treatment effect measure, such as hazard ratios or survival time ratios for time-to-event outcomes (or odds ratios for binary outcomes), population-average conditional treatment effects $d_{ab(P)}$ and population-average marginal treatment effects $\Delta_{ab(P)}(t)$ are not equal and have different interpretations \parencite{Daniel2021,Kahan2014}.
Most notably, the population-average marginal treatment effects $\Delta_{ab(P)}(t)$ vary over time, and depend on the distribution of all prognostic factors, effect modifiers, and baseline hazard in population $P$.
The population-average conditional effects $d_{ab(P)}$ are constant over time and do not depend on the distribution of prognostic factors or baseline hazard in population $P$.
Moreover, different population adjustment methods target different estimands.
MAIC, and STC based on simulation or G-computation, can only produce marginal estimates.
STC based on plugging in mean covariate values is biased for both estimands, and targets neither a conditional or marginal estimand correctly.
Network meta-interpolation produces only conditional estimates, and furthermore cannot typically produce absolute estimates (e.g.\ survival curves or any derivative quantities) which are often required in a decision-making setting.
At present, ML-NMR is the only population-adjustment method that can produce both conditional and marginal estimates, as well as absolute estimates, depending on the requirements for decision-making.

\textcite{Leahy2019} analysed the newly diagnosed multiple myeloma example using multiple MAIC analyses followed by Bayesian NMA.
The inherent limitations of such types of analyses have been described previously \parencite{TSD18}.
In particular, when there are multiple AgD studies, a choice must first be made over which AgD study population to match to.
Then, combining the network of MAIC-adjusted studies and AgD studies in a NMA requires an assumption of constancy of relative effects (i.e.\ that there are no effect modifiers in imbalance between these different populations), which is precisely the assumption that a population-adjusted analysis seeks to relax.
Finally, the resulting estimates are only applicable in a population defined as some weighted average of the included AgD study populations, which may not represent the decision target population.
The ML-NMR analysis addresses each of these issues: it coherently combines evidence from the IPD and AgD studies, accounting for differences between the populations of each study including the AgD studies, and can produce estimates in any target population for decision-making.

A fifth covariate (cytogenetics) was also considered by \textcite{Leahy2019} in some analyses, but we did not include this as it was not measured or reported in all studies. 
Leahy and Walsh found only a very small effect-modifying interaction estimate for this covariate, indicating that potential bias due to omitting cytogenetic factors from the adjustment may be small.
However, this contrasts with subgroup analyses reported by some included trials where more substantial effect modification was observed.
The inability of current population-adjustment methods including ML-NMR to handle different sets of measured or reported covariates across studies is a limitation, and motivates ongoing work that attempts to address this issue.

Both example analyses that we have considered focused on scenarios where event/censoring times were available from each individual in the aggregate studies, for example by reconstructing these from Kaplan-Meier plots \parencite{Guyot2012}.
If individual event/censoring times are not available but instead only conditional log hazard ratios are reported (or log survival time ratios for accelerated failure time models), these may be synthesised directly using a Normal likelihood.
For example, for the conditional log HR of treatment $b$ vs.\ treatment $a$ in study $j$ the likelihood would be $\Normal( \eta_{jb}(\bm{x}^*_j) - \eta_{ja}(\bm{x}^*_j), s^2_{jab} )$,
where $s^2_{jab}$ is the variance of the log HR (given as data) and $\bm{x}^*_j$ is the vector of covariates at the reference levels used in study $j$.
If $\bm{\beta}^*_j=\bm{0}$ then the likelihood simplifies to $\Normal(\gamma_b-\gamma_a, s^2_{jab})$.
Studies with three or more arms would require the correlations between log HRs to be accounted for in the likelihood \parencite{TSD2}.
The limitation of this approach is that it requires the reported log hazard ratios be adjusted in the same manner as the rest of the ML-NMR model.
In theory, it should be possible to instead synthesise reported marginal summary outcomes such as marginal median survival times or marginal (restricted) mean survival times by application of the relationships in equations \cref{eqn:pop_avg_quantiles,eqn:pop_avg_rmst}.
This remains an area for further research.

We have only considered adjusting for covariates measured at baseline: time-varying covariates were not considered since it is likely that, in the aggregate studies, summary covariate information is available only available at baseline and not throughout follow-up.
The inclusion of time-varying covariates in a survival model is often an attempt to correct for observed non-proportionality (i.e.\ failure of the proportional hazards or accelerated failure time assumption).
However, as noted by \textcite[][Section 6.6]{Therneau2000}, such problems may be symptomatic of other issues such as omitted covariates, an incorrect functional form for a covariate, or using an inappropriate model form (e.g.\ a proportional hazards model when an accelerated failure time model would be more appropriate).
Notably, the solutions for these issues can be dealt with within the ML-NMR framework we have described, without requiring further information on time-varying covariates.
Indeed, in the newly diagnosed multiple myeloma example, we found evidence for non-proportional hazards in one study when fitting an unadjusted NMA, but adjusting for baseline covariates in the ML-NMR analysis was sufficient to remove this.

Stratifying the baseline hazard by study is imperative for respecting randomisation within studies, in the same way that we must stratify the intercepts $\mu_j$ by study in the linear predictor.
For example, this is achieved by using study-specific shape parameters $\nu_j$ in a Weibull model or study-specific spline coefficients $\bm{\alpha}_j$ in an M-spline model; all the models detailed in \cref{sec:survival_hazard_functions} are written to stratify baseline hazards by study in this manner.
In this paper, we considered further stratifying the baseline hazard by treatment arm as a way to detect non-proportionality.
If non-proportionality is still present after covariate adjustment, however, the model with baseline hazards stratified by study and treatment arm is not useful for prediction of absolute effects, since survival curves (and all the ensuing summaries) cannot be produced on every treatment in a given population unless all treatment arms have already been observed in that population.
Instead, the models considered here can be extended to incorporate a regression model on the shape of the baseline hazard.
For example, we may include covariate and treatment effects on the shape parameter in a Weibull model via a log link function as $\log(\nu) = \mu^{(\nu)}_j + \bm{x}_{ijk}\tr (\bm{\beta}^{(\nu)}_1 + \bm{\beta}^{(\nu)}_{2,k}) + \gamma^{(\nu)}_k$.
This opens up a further rich and flexible class of models, where departures from non-proportionality can be modelled and absolute predictions can once again be made for any treatment in any population.
Such models are already implemented in the \textit{multinma} R package \parencite{multinma}, where all the available survival models can be specified to include a regression on the shape of the baseline hazard (e.g.\ the Weibull shape parameter or the spline coefficients for the M-spline model).
We leave the description and derivation of these models for a following paper.

When no covariates are available or included in the model, ML-NMR reduces to standard NMA of aggregate data \parencite{Phillippo2020_methods}.
Without covariates, fitting a proportional hazards or accelerated failure time NMA model to the Kaplan-Meier data is essentially equivalent to a simple contrast-based NMA of the log hazard ratios or log survival time ratios with a Normal likelihood (also available in \textit{multinma} \parencite{multinma}).
There is then little benefit to the NMA of having the Kaplan-Meier data available if only proportional hazards or accelerated failure time analyses are to be conducted.
However, having the Kaplan-Meier data available does allow the proportionality assumption to be assessed, and if necessary a model that accounts for departures from proportionality can be fitted even without covariates (as described in the previous paragraph).
Including individual-level covariates in a ML-NMR model can further explain non-proportionality, since the presence of covariate effects means mathematically that the marginal hazard ratios (or marginal survival time ratios) are time-varying and non-proportional.

For the newly diagnosed multiple myeloma example we used M-splines to provide a flexible model on the baseline hazard, which has the attractive property that piecewise exponential models are a special case.
We used a novel random walk prior distribution for the inverse-softmax transformed spline coefficients, which controls the level of smoothing and avoids overfitting through shrinkage.
This approach has several advantages over previous applications of M-splines proposed in other contexts.
\textcite{Brilleman2020} used a Dirichlet prior directly on the spline coefficients, but this does not induce any smoothing or shrinkage and requires careful selection of the number and position of the knots.
\textcite{Jackson_survextrap} used a random effect on the inverse-softmax transformed spline coefficients, centred around a constant baseline hazard, aiming to induce shrinkage and avoid overfitting; however, we found that in practice this did not achieve sufficient shrinkage, with the model complexity and ``wiggliness'' continuing to increase as the number of knots increased, leading to overfitting.
In contrast, our random walk prior distribution does induce sufficient shrinkage to avoid overfitting, as demonstrated in the example, allowing the analyst to simply choose a ``large enough'' number of knots and allow the model to shrink to an appropriate complexity based on the data.
\textcite{Li2022} proposed Bayesian P-splines using a weighted (zero mean) random walk to allow for unevenly-spaced knots and make the prior invariant to knot positioning; we further normalised the knots to make the prior invariant to the number of knots and timescale as well.
This greatly simplifies specification of a hyperprior for the random walk standard deviation, since the scale no longer depends on the number of knots or the timescale, and ensures that unevenly-spaced knots do not affect smoothing or shrinkage behaviour.

Popular alternative flexible models include the Royston-Parmar model \parencite{Freeman2017,Royston2002} and fractional polynomials \parencite{Jansen2011}.
We did not pursue these approaches here for a number of reasons.
Firstly, the Royston-Parmar model places a restricted cubic spline on the log baseline cumulative hazard (or odds), but this is not constrained to be monotonically increasing \parencite{Royston2002}.
This is typically of little consequence in a frequentist setting where sufficient data will ensure that the maximum likelihood estimate lies within the plausible region, but this causes difficulties for a Bayesian analysis that must consider the entire parameter space including the implausible regions.
The fractional polynomial approach models the log hazard as a polynomial function of time and/or log time, with a multivariate treatment effect allowing for non-proportional hazards \parencite{Jansen2011}.
Whilst this is a very flexible approach, the corresponding survival function has no closed form.
Implementation within the ML-NMR framework would therefore require the use of quadrature using the relationship $S(t) = \exp( -\int^t_0 h(u) \dd u )$.
Quadrature has been previously used by other authors to implement alternative forms of flexible models or to allow for time-varying effects \parencite{Brilleman2020,Crowther2014}, and the ML-NMR approach could also be extended in this manner.

Extending the ML-NMR framework to general likelihoods greatly increases the applicability of this approach, including to the very common scenario of population adjustment for survival outcomes.
The Stan code that we have developed and provided in the supplementary materials is modular, and all that is required to fit a range of alternative models in the ML-NMR framework is to specify the form of the survival and hazard functions for the individual-level model.
Once these have been specified, the numerical integration step to obtain the individual marginal likelihood remains the same, and is automatically implemented in the Stan code.
Whilst not described here, it is also straightforward to account for left censoring, interval censoring, and left truncation (delayed entry) in this framework in the standard manner by considering the appropriate contributions from the survival function (e.g. as summarised by \textcite{Brilleman2020}), and all of these are implemented in the \textit{multinma} R package \parencite{multinma}.
The \textit{multinma} R package provides a user-friendly interface to implementing ML-NMR, AgD NMA, and IPD NMA models for a wide range of data types, supporting the uptake of these methods by analysts in practical applications.

\if\blind0
\section*{Acknowledgements}
This work was supported by the UK Medical Research Council, grant numbers MR/P015298/1, MR/R025223/1, and MR/W016648/1.
\fi

\printbibliography[heading=bibintoc,title={Bibliography}]

\clearpage
\appendix
\setcounter{secnumdepth}{3}
\setcounter{equation}{0}
\setcounter{table}{0}
\setcounter{figure}{0}
\renewcommand{\theequation}{\Alph{section}.\arabic{equation}}
\renewcommand{\thetable}{\Alph{section}.\arabic{table}}
\renewcommand{\thefigure}{\Alph{section}.\arabic{figure}}
\pagenumbering{arabic}

\section{Implications for binary outcome data}\label{ssec:gl_binary_derivation}
Suppose that we have binary outcomes $y_{ijk} \sim \Bernoulli(p_{ijk})$.
In this case, the individual conditional likelihood contributions are
\begin{equation*}
  L^\mathrm{Con}_{ijk | \bm{x}}(\bm{\xi} ; y_{ijk}, \bm{x}_{ijk}) = p_{ijk}^{y_{ijk}} (1 - p_{ijk})^{(1 - y_{ijk})},
\end{equation*}
where the individual event probabilities $p_{ijk}$ are modelled using $p_{ijk} = g^{-1}(\eta_{jk}(\bm{x}_{ijk}))$ with a suitable link function $g(\cdot)$ (e.g.\ a logit or probit link function).
Using equation \eqref{eqn:gl_individual_marg_lik}, the individual marginal likelihood contributions are
\begin{align*}
  L^\mathrm{Mar}_{ijk}(\bm{\xi} ; y_{ijk}) &= \int_{\mathfrak{X}} L^\mathrm{Con}_{ijk | \bm{x}}(\bm{\xi} ; y_{ijk}, \bm{x}) f_{jk}(\bm{x}) \dd\bm{x} \\
  &= \int_{\mathfrak{X}} g^{-1}(\eta_{jk}(\bm{x}))^{y_{ijk}} \left(1 - g^{-1}(\eta_{jk}(\bm{x})) \right)^{(1 - y_{ijk})} f_{jk}(\bm{x}) \dd\bm{x} \\
  &= \bar{p}_{jk}^{y_{ijk}} (1 - \bar{p}_{jk})^{(1 - y_{ijk})}
\end{align*}
where $\bar{p}_{jk} = \int_{\mathfrak{X}} g^{-1}(\eta_{jk}(\bm{x})) f_{jk}(\bm{x}) \dd\bm{x}$ is the mean event probability on treatment $k$ in study $j$, since $y_{ijk} \in \{ 0,1 \}$.
The aggregate likelihood contribution for $y_{\bullet jk}$ events out of $N_{jk}$ individuals on treatment $k$ in study $j$ is then proportional to the product of $y_{\bullet jk}$ many $L^\mathrm{Mar}_{ijk}(\bm{\xi}; 1)$ terms and $N_{jk} - y_{\bullet jk}$ many $L^\mathrm{Mar}_{ijk}(\bm{\xi}; 0)$ terms:
\begin{align*}
  L^\mathrm{Mar}_{\bullet jk}(\bm{\xi} ; y_{\bullet jk}) &\propto L^\mathrm{Mar}_{ijk}(\bm{\xi}; 1)^{y_{\bullet jk}} L^\mathrm{Mar}_{ijk}(\bm{\xi}; 0)^{(N_{jk} - y_{\bullet jk})} \\
  &= \bar{p}_{jk}^{y_{\bullet jk}} (1 - \bar{p}_{jk})^{(N_{jk} - y_{\bullet jk})}
\end{align*}
which we recognise as a $\Binomial(N_{jk}, \bar{p}_{jk})$ likelihood.
In other words, we recover the one-parameter Binomial likelihood described by \textcite{Phillippo2020_methods}.

\textcite{Phillippo2020_methods} improved upon the one-parameter Binomial likelihood with a two-parameter Binomial likelihood in which both $\bar{p}_{jk}$ and $N_{jk}$ were adjusted, aiming to obtain a likelihood closer to the ``true'' Poisson Binomial aggregate likelihood.
The Poisson Binomial likelihood describes the total number of events given a vector of individual probabilities, where the exact individuals experiencing an event are unknown; however, we cannot use this likelihood directly as the parameter vector is not identifiable given the aggregate data.
Instead, the one-parameter Binomial likelihood assigns the same event probability $\bar{p}_{jk}$ to each individual on treatment $k$ in study $j$; however, this is not the most efficient model since we know that the individual event probabilities differ.
The two-parameter Binomial likelihood acknowledges this, and as a result has a smaller variance (matching that of the Poisson Binomial).
Given full IPD, the individual-level Bernoulli likelihood would additionally make use of the information on precisely which individuals experienced events.
Intuitively then, the two-parameter Binomial likelihood lies in between the one-parameter Binomial likelihood and the full IPD individual-level Bernoulli likelihood in terms of efficiency.
The marginal likelihood approach is not ``wrong'' here, it is just not the most efficient.
In this case, we can improve on the one-parameter Binomial likelihood obtained through the marginal likelihood approach since we know the ``true'' form of the aggregate likelihood, although in practice we find that this makes little difference to the results.
However, this will not be possible in general, as we cannot always derive (or even approximate) the appropriate likelihood distribution for the aggregate data.

\setcounter{equation}{0}
\setcounter{table}{0}
\setcounter{figure}{0}
\section{Survival and hazard functions for models implemented in \textit{multinma}}\label{sec:survival_hazard_functions}
The \textit{multinma} R package \parencite{multinma} implements a range of models for survival outcomes.
We detail these here, with their survival and hazard functions $S_{jk}(t|\bm{x})$ and $h_{jk}(t|\bm{x})$.

\subsection{Proportional hazards models}\label{sec:ph_models}
For all proportional hazards models described below, the linear predictor $\eta_{jk}(\bm{x})$ is placed on the log hazard scale.
Model coefficients are interpreted as log hazard ratios, with positive values representing an increased hazard of experiencing the event.
For example, a coefficient of $\log(2)$ represents a doubling of the hazard of experiencing the event.

\subsubsection{Exponential (proportional hazards form)}

An Exponential proportional hazards model has survival and hazard functions
\begin{subequations}
\begin{align}
	S_{jk}(t|\bm{x}) &= \exp\mleft( -t \exp(\eta_{jk}(\bm{x})) \mright)\\
	h_{jk}(t|\bm{x}) &= \exp(\eta_{jk}(\bm{x}))
\end{align}
\end{subequations}

\subsubsection{Weibull (proportional hazards form)}

A Weibull proportional hazards model has survival and hazard functions
\begin{subequations}
\begin{align}
	S_{jk}(t|\bm{x}) &= \exp\mleft( -t^{\nu_j} \exp(\eta_{jk}(\bm{x})) \mright)\\
	h_{jk}(t|\bm{x}) &= \nu_j  t^{\nu_j - 1} \exp(\eta_{jk}(\bm{x}))
\end{align}
\end{subequations}
where $\nu_j > 0$ is a study-specific shape parameter.
The shape parameters $\nu_j$ are given prior distributions; the default in the \textit{multinma} package is a weakly-informative prior distribution $\nu_j \sim \halfNormal(0, 10^2)$.

\subsubsection{Gompertz}

A Gompertz proportional hazards model has survival and hazard functions
\begin{subequations}
\begin{align}
	S_{jk}(t|\bm{x}) &= \exp\mleft( -\nu_j^{-1} \exp(\eta_{jk}(\bm{x})) (\exp(t\nu_j) - 1) \mright) \\
	h_{jk}(t|\bm{x}) &= \exp\mleft( \eta_{jk}(\bm{x}) \mright) \exp(t\nu_j)
\end{align}
\end{subequations}
where $\nu_j > 0$ is a study-specific scale parameter.
The scale parameters $\nu_j$ are given prior distributions; the default in the \textit{multinma} package is a weakly-informative prior distribution $\nu_j \sim \halfNormal(0, 10^2)$.

\subsubsection{M-spline}\label{ssec:mspline_model}

We propose a flexible model with M-splines on the baseline hazard based on the M-spline model of \textcite{Brilleman2020}, with survival and hazard functions
\begin{subequations}
\begin{align}
	S_{jk}(t|\bm{x}) &= \exp\mleft( -\bm{\alpha}_j\tr \bm{I}(t, \bm{\zeta}_j) \exp(\eta_{jk}(\bm{x})) \mright)\\
	h_{jk}(t|\bm{x}) &= \bm{\alpha}_j\tr \bm{M}(t, \bm{\zeta}_j) \exp(\eta_{jk}(\bm{x}))
\end{align}
\end{subequations}
where $\bm{\alpha}_j$ is a study-specific vector of spline coefficients, $\bm{M}_\kappa(t, \bm{\zeta}_j)$ is the M-spline basis of order $\kappa$ with a study-specific knot sequence $\bm{\zeta}_j$ evaluated at time $t$, and $\bm{I}_\kappa(t, \bm{\zeta}_j)$ is the corresponding integrated M-spline basis (an I-spline basis).
The basis polynomials have degree $\kappa-1$, so a basis of order 4 corresponds to a cubic M-spline basis.

The knot sequence $\bm{\zeta}_j = (\zeta_{j,0}, \cdots, \zeta_{j, L+1})$ is a strictly increasing vector of length $L+2$, where $L$ is the number of internal knots chosen by the user, and $\zeta_{j,0}$ and $\zeta_{j,L+1}$ are the lower and upper boundary knots.
Internal knots may be placed at arbitrary locations within the boundary knots; by default, we choose the place these at evenly-spaced quantiles of the observed event times in each study, and place boundary knots at time 0 and the last event/censoring time in each study.
The dimension of the spline basis, i.e. the number of spline coefficients in the vector $\bm{\alpha}_j$, is equal to $L+\kappa$.

The M-spline and I-spline bases are constructed using the recursive formulae of \textcite{Ramsay1988}, which are implemented in the \textit{splines2} R package \parencite{splines2}.
Following \textcite{Ramsay1988}, define an augmented knot vector $\bm{\zeta}_j^*$ of length $L+2\kappa$ by 
\begin{equation}
	\zeta^*_{j,s} = 
	\begin{cases}
		\zeta_{j,0} & \textrm{for } s = 1, \dots, \kappa \\
		\zeta_{j,s-\kappa} & \textrm{for } s = \kappa+1, \dots, \kappa+L \\
		\zeta_{j,L+1} & \textrm{for } s = \kappa+L+1, \dots, L+2\kappa
	\end{cases}
\end{equation}
In other words, we pad the knot vector $\bm{\zeta}_j$ by $\kappa$ replications of the lower and upper boundary knots at the start and end of the vector respectively to obtain the augmented knot vector $\bm{\zeta}^*_j$.
Then the M-spline basis $M_{\kappa, s}(t, \bm{\zeta}^*_j)$ at a given time $t$ for each dimension $s = 1,\dots,L+\kappa$ is defined recursively by
\begin{subequations}
\begin{align}
	M_{1, s}(t, \bm{\zeta}^*_j) &= \frac{1}{\zeta^*_{j, s+1} - \zeta^*_{j, s}} \quad \textrm{if } \zeta^*_{j, s} \le t < \zeta^*_{j, s+1} \textrm{, otherwise } 0 \\
  M_{r, s}(t, \bm{\zeta}^*_j) &= \frac{r \mleft( (t - \zeta^*_{j,s}) M_{r-1, s}(t, \bm{\zeta}^*_j) + (\zeta^*_{j,s+1} - t) M_{r-1, s+1}(t, \bm{\zeta}^*_j) \mright)}{(r-1)(\zeta^*_{j, s+r} - \zeta^*_{j, s})} \\
\end{align}
\end{subequations}
for $r = 2,\dots,\kappa$.
We note two particular properties of this M-spline basis.
Firstly, $M_{\kappa,s}(t, \bm{\zeta}^*_j) > 0$ only for $\zeta^*_{j, s} \le t < \zeta^*_{j, s+\kappa}$, and is zero everywhere else. 
In other words, the spline fit between knots $\zeta_{j,s}$ and $\zeta_{j,s+1}$ is only informed by data up to $\kappa-1$ knots later, up to time $\zeta_{j,s+\kappa}$. 
Secondly, $\int_{\zeta_{j,0}}^{\zeta_{j,L+1}} \sum_{s=1}^{L+\kappa} M_{\kappa,s}(t,\bm{\zeta}^*_j) \dd t =1$, that is the M-spline is normalised to have integral 1 between the boundary knots.

The corresponding I-spline basis evaluated at time $t$ is the integral of the M-spline basis from the lower boundary point up until time $t$:
\begin{equation}
	I_{\kappa,s}(t, \bm{\zeta}_j) = \int_{\zeta_{j,0}}^t M_{\kappa,s}(v, \bm{\zeta}_j) \dd v.
\end{equation}

The spline coefficients $\bm{\alpha}_j$ lie in the unit simplex, i.e.\ $0 \le \alpha_{j,s} \le 1$ $\forall s = 1,\dots,L+\kappa$, and $\sum_{s=1}^{L+\kappa} \alpha_{j,s}=1$.
To avoid overfitting, we propose the use of a random walk prior distribution on the inverse-softmax transformed spline coefficients.
In practice, this means that the analyst simply needs to specify a sufficiently-large number of knots (we choose 7 internal knots as the default in \textit{multinma}), which are then smoothed over time and shrunk towards a constant baseline hazard.
The amount of smoothing and shrinkage is controlled by the standard deviation of the random walk, which is given a prior distribution and estimated from the data.
The random walk prior distribution is defined as follows:
\begin{subequations}
\begin{align}
	\bm{\alpha}_j &= \operatorname{softmax}(\bm{\alpha_j^*}) \\
	\alpha^*_{j,l} &= c_{j,l} + \sum_{m = 1}^l u_{j,m} \quad \forall l=1,\dots,L+\kappa-1\\
	u_{j,l} &\sim \Normal(0, \sigma^2_j w_{j,l}) \quad \forall l=1,\dots,L+\kappa-1
\end{align}
\end{subequations}
where the softmax function maps a real-valued vector $\bm{\alpha_j^*}$ of length $L+\kappa-1$ to the coefficient vector $\bm{\alpha}_j$ of length $L+\kappa$ on the unit simplex:
\begin{equation}
	\operatorname{softmax}(\bm{\alpha}^*_j) = \frac{\mleft\lbrack 1, \exp(\bm{\alpha}^*_j)\tr \mright\rbrack\tr}{1 + \sum_{l=1}^{L+\kappa-1} \exp(\alpha^*_{j,l})}
\end{equation}
with inverse
\begin{equation}
	\operatorname{softmax}^{-1}(\bm{\alpha}_j) = \log(\bm{\alpha}_{j,2:(L+\kappa)}) - \log(\alpha_{j,1}).
\end{equation}
Rather than having zero mean, this random walk is centred around a prior mean vector $\bm{c}_j$ which corresponds to a constant baseline hazard.
This idea borrows from \textcite{Jackson_survextrap}, who used a random effect prior on $\bm{\alpha}_j$ with mean $\bm{c}_j$, deriving $\bm{c}_j$ from the augmented knot vector $\bm{\zeta}^*_j$ as 
\begin{equation}
	\bm{c}_j = \operatorname{softmax}^{-1}\mleft( \frac{\bm{\zeta}^*_{j,\kappa:(L+2\kappa)} - \bm{\zeta}^*_{j,1:(L+\kappa)}}{\kappa (\zeta_{j,L+1} - \zeta_{j,0})} \mright).
\end{equation}
The random walk standard deviation $\sigma_j$ controls the amount of smoothing and shrinkage; as $\sigma_j$ approaches zero the baseline hazard becomes smoother (less ``wiggly'') and approaches a constant baseline hazard.
Furthermore, we introduce weights $\bm{w}_j$ into the random walk which are defined by the distance between each pair of knots:
\begin{equation}
	\bm{w}_j = \frac{\bm{\zeta}^*_{j,(\kappa+1):(L+2\kappa-1)} - \bm{\zeta}^*_{j,2:(L+\kappa)}}{(\kappa-1)(\zeta_{j,L+1} - \zeta_{j,0})}.
\end{equation}
This follows a similar approach to the Bayesian P-splines proposed by \textcite{Li2022}, except that here we additionally normalise the weights to sum to 1.
These weights serve two purposes.
Firstly, the weights allow the overall prior distribution on the baseline hazard to be invariant to the knot locations $\bm{\zeta}_j$, even when these are unevenly spaced over time.
Without the weights, time periods with a greater number of knots (typically the start of follow-up, when we choose the knot locations by quantiles of observed survival times) will have greater prior variation, resulting in under-smoothing \parencite{Li2022}.
Secondly, normalising the weights controls the overall variation in the prior, making the overall prior distribution on the baseline hazard invariant to the number of knots and to the timescale (i.e.\ $\zeta_{j,L+1} - \zeta_{j,0}$).
Without normalising, the overall variation in the prior increases as the number of knots increases or as the timescale decreases; increasing the number of knots or decreasing the timescale (e.g.\ counting time in weeks instead of days) would require a tighter hyperprior distribution on $\sigma_j$ to achieve the same level of smoothing. 
These weights greatly simplify specification of a hyperprior distribution for the random walk standard deviation $\sigma_j$, as its interpretation no longer depends on the number or spacing of the knots, or on the timescale.
We choose $\sigma_j \sim \halfNormal(0, 1^2)$ as a weakly-informative prior distribution.

\subsubsection{Piecewise Exponential}\label{ssec:pexp_model}
The piecewise exponential model is a special case of the M-spline model, where the degree of the M-spline basis is 0 (i.e.\ $\kappa = 1$).
In this case, the hazard will be constant within each pair of knots.
We choose to retain an identical setup to the M-spline model, including the use of a random walk prior distribution on the (inverse-softmax transformed) spline coefficients to allow smoothing over time and shrinkage towards a constant baseline hazard.

\subsection{Accelerated failure time models}\label{sec:aft_models}

For all accelerated failure time models described below, the linear predictor $\eta_{jk}(\bm{x})$ is placed on the log time scale.
We parameterise all the accelerated failure time models listed here such that the model coefficients are interpreted as log survival time ratios (equal to minus log acceleration factors), with positive values representing an increased expected survival time.
For example, a coefficient of $\log(2)$ represents a doubling of the expected survival time.

\subsubsection{Exponential (accelerated failure time form)}

An Exponential accelerated failure time model has survival and hazard functions
\begin{subequations}
\begin{align}
	S_{jk}(t|\bm{x}) &= \exp\mleft( -t \exp(-\eta_{jk}(\bm{x})) \mright)\\
	h_{jk}(t|\bm{x}) &= \exp(-\eta_{jk}(\bm{x}))
\end{align}
\end{subequations}

\subsubsection{Weibull (accelerated failure time form)}

A Weibull accelerated failure time model has survival and hazard functions
\begin{subequations}
\begin{align}
	S_{jk}(t|\bm{x}) &= \exp\mleft( -t^{\nu_j} \exp(-\nu_j \eta_{jk}(\bm{x})) \mright)\\
	h_{jk}(t|\bm{x}) &= \nu_j  t^{\nu_j - 1} \exp(-\nu_j \eta_{jk}(\bm{x}))
\end{align}
\end{subequations}
where $\nu_j > 0$ is a study-specific shape parameter.
The shape parameters $\nu_j$ are given prior distributions; the default in the \textit{multinma} package is a weakly-informative prior distribution $\nu_j \sim \halfNormal(0, 10^2)$.

\subsubsection{log-Normal}

A log-Normal accelerated failure time model has survival and hazard functions
\begin{subequations}
\begin{align}
	S_{jk}(t|\bm{x}) &= 1 - \operatorname{\Phi}\mleft( \frac{\log(t) - \eta_{jk}(\bm{x})}{\sigma_j} \mright)\\
	h_{jk}(t|\bm{x}) &= \frac{(t\sigma_j)^{-1} \operatorname{\phi}\mleft( \frac{\log(t) - \eta_{jk}(\bm{x})}{\sigma_j} \mright)}{1 - \operatorname{\Phi}\mleft( \frac{\log(t) - \eta_{jk}(\bm{x})}{\sigma_j} \mright)}
\end{align}
\end{subequations}
where $\sigma_j > 0$ is a study-specific log-scale parameter, and $\mathop{\phi}(\cdot)$ and $\mathop{\Phi}(\cdot)$ are the standard Normal density and cumulative distribution functions.
The log-scale parameters $\sigma_j$ are given prior distributions; the default in the \textit{multinma} package is a weakly-informative prior distribution $\sigma_j \sim \halfNormal(0, 10^2)$.

\subsubsection{log-Logistic}

A log-Logistic accelerated failure time model has survival and hazard functions
\begin{subequations}
\begin{align}
	S_{jk}(t|\bm{x}) &= \frac{1}{1 + \mleft( t\exp(-\eta_{jk}(\bm{x})) \mright)^{\nu_j}}\\
	h_{jk}(t|\bm{x}) &= \frac{\nu_j\exp(-\eta_{jk}(\bm{x})) \mleft( t\exp(-\eta_{jk}(\bm{x})) \mright)^{\nu_j-1}}{1 + \mleft( t\exp(-\eta_{jk}(\bm{x})) \mright)^{\nu_j}}
\end{align}
\end{subequations}
where $\nu_j > 0$ is a study-specific shape parameter.
The shape parameters $\nu_j$ are given prior distributions; the default in the \textit{multinma} package is a weakly-informative prior distribution $\nu_j \sim \halfNormal(0, 10^2)$.

\subsubsection{Gamma}

A Gamma accelerated failure time model has survival and hazard functions
\begin{subequations}
\begin{align}
	S_{jk}(t|\bm{x}) &= 1 - \frac{\mathop{\gamma}\mleft( \nu_j, t\exp(-\eta_{jk}(\bm{x})) \mright)}{\mathop{\Gamma}(\nu_j)} \\
	h_{jk}(t|\bm{x}) &= \frac{t^{\nu_j-1} \exp(-t \exp(-\eta_{jk}(\bm{x}))) \exp(-\eta_{jk}(\bm{x}))^{\nu_j}}{\mathop{\Gamma}(\nu_j) - \mathop{\gamma}\mleft( \nu_j, t\exp(-\eta_{jk}(\bm{x})) \mright)}
\end{align}
\end{subequations}
where $\nu_j > 0$ is a study-specific shape parameter, $\mathop{\Gamma(\cdot)}$ is the Gamma function, and $\mathop{\gamma}(\cdot, \cdot)$ is the incomplete Gamma function.
The shape parameters $\nu_j$ are given prior distributions; the default in the \textit{multinma} package is a weakly-informative prior distribution $\nu_j \sim \halfNormal(0, 10^2)$.

\subsubsection{Generalised Gamma}

The generalised Gamma distribution permits a range of behaviours for the baseline hazard including increasing, decreasing, bathtub and arc-shaped hazards.
We define a generalised Gamma accelerated failure time model following the parameterisation of \textcite{Lawless1980}, which has survival and hazard functions
\begin{subequations}
	\begin{align}
		S_{jk}(t|\bm{x}) &= 1 - \frac{\mathop{\gamma}\mleft( \nu_j, \nu_j^{-0.5} t\exp\mleft(-\eta_{jk}(\bm{x}) \sigma_j^{-1}\mright) \mright)}{\mathop{\Gamma}(\nu_j)} \\
		h_{jk}(t|\bm{x}) &= \frac{\nu_j^{-0.5} \mleft( \nu_j (t\exp(-\eta_{jk}(\bm{x})))^{\frac{\nu_j^{-0.5}}{\sigma_j}} \mright)^{\nu_j} \exp\mleft( -\nu_j (t\exp(-\eta_{jk}(\bm{x})))^{\frac{\nu_j^{-0.5}}{\sigma_j}} \mright)}{t\sigma_j  \mleft( \mathop{\Gamma}(\nu_j) - \mathop{\gamma}\mleft( \nu_j, \nu_j^{-0.5} t\exp\mleft(-\eta_{jk}(\bm{x}) \sigma_j^{-1}\mright) \mright) \mright)}
	\end{align}
	\end{subequations}
where $\sigma_j > 0$ is a study-specific scale parameter, $\nu_j > 0$ is a study-specific shape parameter $\mathop{\Gamma(\cdot)}$ is the Gamma function, and $\mathop{\gamma}(\cdot, \cdot)$ is the incomplete Gamma function.
The scale and shape parameters are given prior distributions; the defaults in the \textit{multinma} package are weakly-informative prior distributions $\sigma_j \sim \halfNormal(0, 10^2)$ and $\nu_j \sim \halfNormal(0, 10^2)$.

This parameterisation is related to that discussed by \textcite{Cox2007} with $Q = \nu^{-0.5}$.
The parameterisation used here effectively bounds the shape parameter $\nu$ away from numerical instabilities as $\nu \rightarrow \infty$ (i.e.\ away from $Q \rightarrow 0$, the log-Normal distribution) with any proper prior distribution on $\nu$.
Implicitly, this parameterisation is restricted to $Q > 0$ and so certain survival distributions like the inverse-Gamma and inverse-Weibull are not part of the parameter space; however, $Q > 0$ still encompasses all the other parametric accelerated failure time models listed here.

\clearpage
\setcounter{equation}{0}
\setcounter{table}{0}
\setcounter{figure}{0}
\section{Simulated survival example}\label{sec:simulated_example_details}

\subsection{Simulated scenario}
We simulate individuals in two studies, an $AB$ study randomising 500 individuals 1:1 to $A$ and $B$, and an $AC$ study randomising 400 individuals 1:1 to $A$ and $C$.
For each individual, we generate three covariates according to the following (independent) distributions in each study:
\begin{align*}
	X_{1(AB)} &\sim \Normal\mleft(0, 0.5^2\mright), & X_{2(AB)} &\sim \Gammadist(4, 2), & X_{3(AB)} &\sim \Bernoulli(0.2) \\
	X_{1(AC)} &\sim \Normal\mleft(1, 0.4^2\mright), & X_{2(AC)} &\sim \Gammadist(6, 2), & X_{3(AC)} &\sim \Bernoulli(0.7)
\end{align*}

\begin{figure}[htbp]
	\centering
	\includegraphics[width=\textwidth]{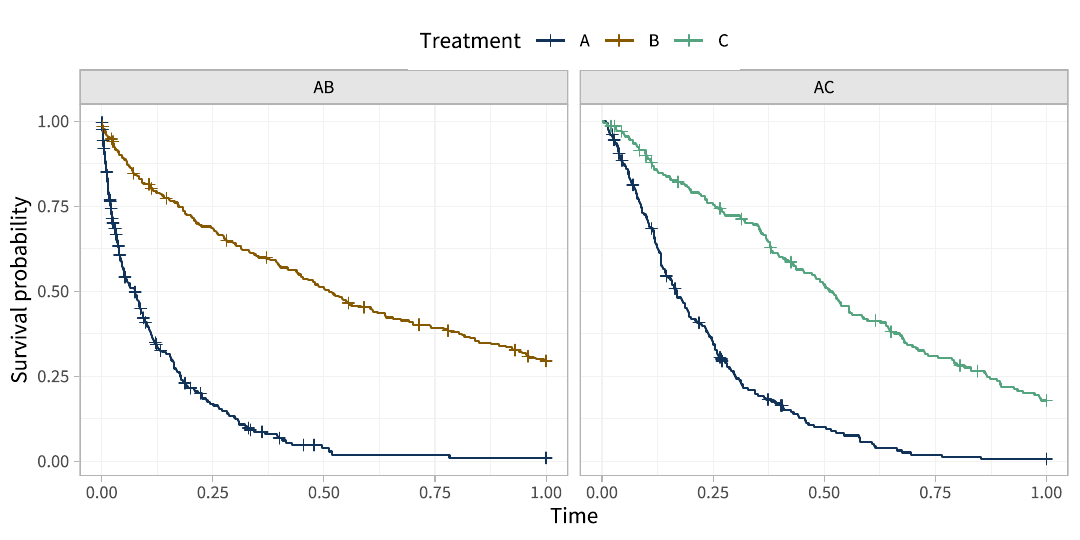}
	\caption{Simulated Kaplan-Meier survival curves for each treatment in each study. Censored events are marked with a cross ($+$).}
	\label{fig:survival_sim_curves}
\end{figure}

We simulate survival times from a Weibull model in each study, with shape parameters $\nu_{(AB)}=0.8$, $\nu_{(AC)}=1.2$ and linear predictor
\begin{equation}\label{eqn:sim_linear_predictor}
	\eta_{jk}(\bm{x}_{ijk}) = \mu_j \bm{x}\tr_{ijk}(\bm{\beta}_1 + \bm{\beta}_{2,k}) + \gamma_k,
\end{equation}
with baseline log rates $\mu_{(AB)} = \log(6.2)$ and $\mu_{(AC)} = \log(5.8)$, prognostic coefficients $\bm{\beta_1} = (0.1, 0.05, -0.25)\tr$, effect modifying coefficients $\bm{\beta}_{2,B} = \bm{\beta}_{2,C} = (-0.2, -0.2, -0.1)\tr$ (and $\bm{\beta}_{2,A} = \bm{0}$), and individual-level treatment effects $\gamma_B = -1.2$, $\gamma_C = -0.5$ and $\gamma_A = 0$.
All covariates are both prognostic and effect modifying, and we have $\bm{\beta}_{2,B} = \bm{\beta}_{2,C}$ so the shared effect modifier assumption holds.
Survival times are simulated using the Cumulative Distribution Function inversion method described by \textcite{Bender2005}, implemented in the R package \textit{simsurv} \parencite{simsurv}.
We censor all surviving individuals at time $t=1$ for both studies, and further uniformly censor 10\% of individuals within each study.
The resulting Kaplan-Meier survival curves are shown in \cref{fig:survival_sim_curves}.
For the ML-NMR analysis, we use only summary covariate information for the $AC$ trial (means and standard deviations for the continuous covariates, and the proportion for the discrete covariate).

\subsection{Additional tables and figures for the simulated survival example}

\begin{table}[htbp]
	\footnotesize
	\centering
	\caption{Model comparison results, using full IPD NMA and ML-NMR. The leave-one-out information criterion (LOOIC) is equal to $-2\cdot\textrm{ELPD}$, where ELPD is the expected log pointwise predictive density, and lower LOOIC values indicate better expected predictive performance. $p_\mathrm{LOO}$ is the effective number of parameters. The ELPD differences are in comparison with the respective Weibull models, with negative values favouring the Weibull model. Standard errors for each statistic are given alongside in small brackets.}

	\label{tab:looic_comparison}
	\begin{tabular}{lr@{\hspace*{0.5\tabcolsep}}>{\tiny}rr@{\hspace*{0.5\tabcolsep}}>{\tiny}rr@{\hspace*{0.5\tabcolsep}}>{\tiny}rr@{\hspace*{0.5\tabcolsep}}>{\tiny}rr@{\hspace*{0.5\tabcolsep}}>{\tiny}rr@{\hspace*{0.5\tabcolsep}}>{\tiny}r}
		\toprule
		& \multicolumn{6}{c}{ML-NMR} & \multicolumn{6}{c}{IPD NMA} \\
		\cmidrule(r){2-7} \cmidrule(l){8-13}
		& \multicolumn{2}{c}{Exponential} & \multicolumn{2}{c}{Gompertz} & \multicolumn{2}{c}{Weibull} & \multicolumn{2}{c}{Exponential} & \multicolumn{2}{c}{Gompertz} & \multicolumn{2}{c}{Weibull} \\
		\midrule
		 LOOIC & $-201.7$ & ($67.2$) & $-215.1$ & ($67.7$) & $-261.8$ & ($69.5$) & $-221.0$ & ($67.5$) & $-237.4$ & ($68.0$) & $-283.7$ & ($69.8$) \\ 
  ELPD & $100.9$ & ($33.6$) & $107.5$ & ($33.9$) & $130.9$ & ($34.7$) & $110.5$ & ($33.8$) & $118.7$ & ($34.0$) & $141.8$ & ($34.9$) \\ 
  $p_\mathrm{LOO}$ & $10.9$ & ($1.1$) & $13.9$ & ($1.3$) & $12.3$ & ($0.9$) & $10.7$ & ($0.9$) & $13.0$ & ($1.3$) & $12.3$ & ($1.0$) \\ 
  ELPD difference & $-30.0$ & ($8.3$) & $-23.3$ & ($7.4$) &  &  & $-31.3$ & ($8.3$) & $-23.1$ & ($7.5$) &  &  \\ 
  
		\bottomrule
	\end{tabular}
\end{table}

\begin{figure}[htbp]
	\centering
	\includegraphics[width=.8\textwidth]{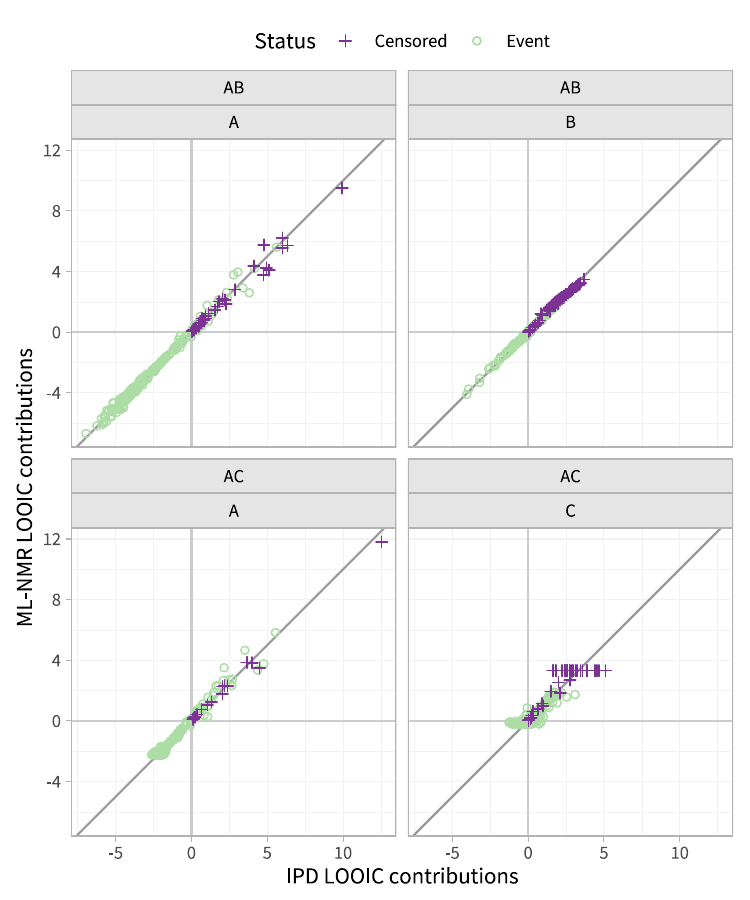}
	\caption{Contributions to the LOOIC from each event and censoring time in the Weibull model, plotted for ML-NMR using only summary covariate information in the $AC$ study against an IPD NMA with full information from every study. The LOOIC contributions follow the straight line of equality well, showing that the same observations are fitted similarly well whether the full IPD was used or aggregate $AC$ data.	The only noticeable exception to this is a horizontal cluster of LOOIC contributions for a set of censoring times in the $C$ treatment arm of the AgD $AC$ trial, which all have LOOIC contributions around 2.5 under the ML-NMR model.	These correspond to individuals all censored at the end of the trial ($t=1$), which under the ML-NMR model are all given the same marginal likelihood contribution.}
	\label{fig:survival_looic_contributions}
\end{figure}

\begin{table}[htbp]
	\footnotesize
	\centering
	\caption[Table of estimated model parameters from the ML-NMR model and the full IPD NMA, alongside the true values used in the simulation]{Table of estimated model parameters and 95\% Credible Intervals from the ML-NMR model and the full IPD NMA, alongside the true values used in the simulation.}
	\label{tab:survival_parameter_ests}
	\begin{tabular}{llccc}
		\toprule
		\multicolumn{2}{c}{Parameter} & Truth & IPD NMA & ML-NMR \\
		\midrule
		 Treatment Effect & $\gamma_B$ & $-1.20$ & $-0.97$ & $-1.20$ \\ 
   &  &  & ($-1.30$, $-0.64$) & ($-1.66$, $-0.72$) \\ 
   [0.5ex] & $\gamma_C$ & $-0.50$ & $-0.10$ & $-0.30$ \\ 
   &  &  & ($-0.71$, $0.47$) & ($-1.10$, $0.53$) \\ 
   [0.5ex]Prognostic Effect & $\beta_{1;1}$ & $0.10$ & $0.15$ & $0.32$ \\ 
   &  &  & ($-0.06$, $0.35$) & ($0.03$, $0.61$) \\ 
   [0.5ex] & $\beta_{1;2}$ & $0.05$ & $0.09$ & $0.00$ \\ 
   &  &  & ($0.00$, $0.17$) & ($-0.14$, $0.14$) \\ 
   [0.5ex] & $\beta_{1;3}$ & $-0.25$ & $-0.23$ & $-0.11$ \\ 
   &  &  & ($-0.47$, $0.02$) & ($-0.48$, $0.23$) \\ 
   [0.5ex]EM Interaction & $\beta_{2;1}$ & $-0.20$ & $-0.24$ & $-0.39$ \\ 
   &  &  & ($-0.53$, $0.06$) & ($-0.79$, $0.01$) \\ 
   [0.5ex] & $\beta_{2;2}$ & $-0.20$ & $-0.26$ & $-0.14$ \\ 
   &  &  & ($-0.39$, $-0.14$) & ($-0.34$, $0.06$) \\ 
   [0.5ex] & $\beta_{2;3}$ & $-0.10$ & $-0.25$ & $-0.27$ \\ 
   &  &  & ($-0.60$, $0.11$) & ($-0.79$, $0.23$) \\ 
   [0.5ex]Shape & $\nu_{AB}$ & $0.80$ & $0.78$ & $0.78$ \\ 
   &  &  & ($0.72$, $0.85$) & ($0.72$, $0.85$) \\ 
   [0.5ex] & $\nu_{AC}$ & $1.20$ & $1.28$ & $1.27$ \\ 
   &  &  & ($1.17$, $1.39$) & ($1.17$, $1.39$) \\ 
  
		\bottomrule
	\end{tabular}
\end{table}

\clearpage
\setcounter{equation}{0}
\setcounter{table}{0}
\setcounter{figure}{0}
\section{Producing synthetic data for the newly diagnosed multiple myeloma example}\label{ssec:synthetic_data}
\begin{figure}[htbp]
  \includegraphics[width=\textwidth]{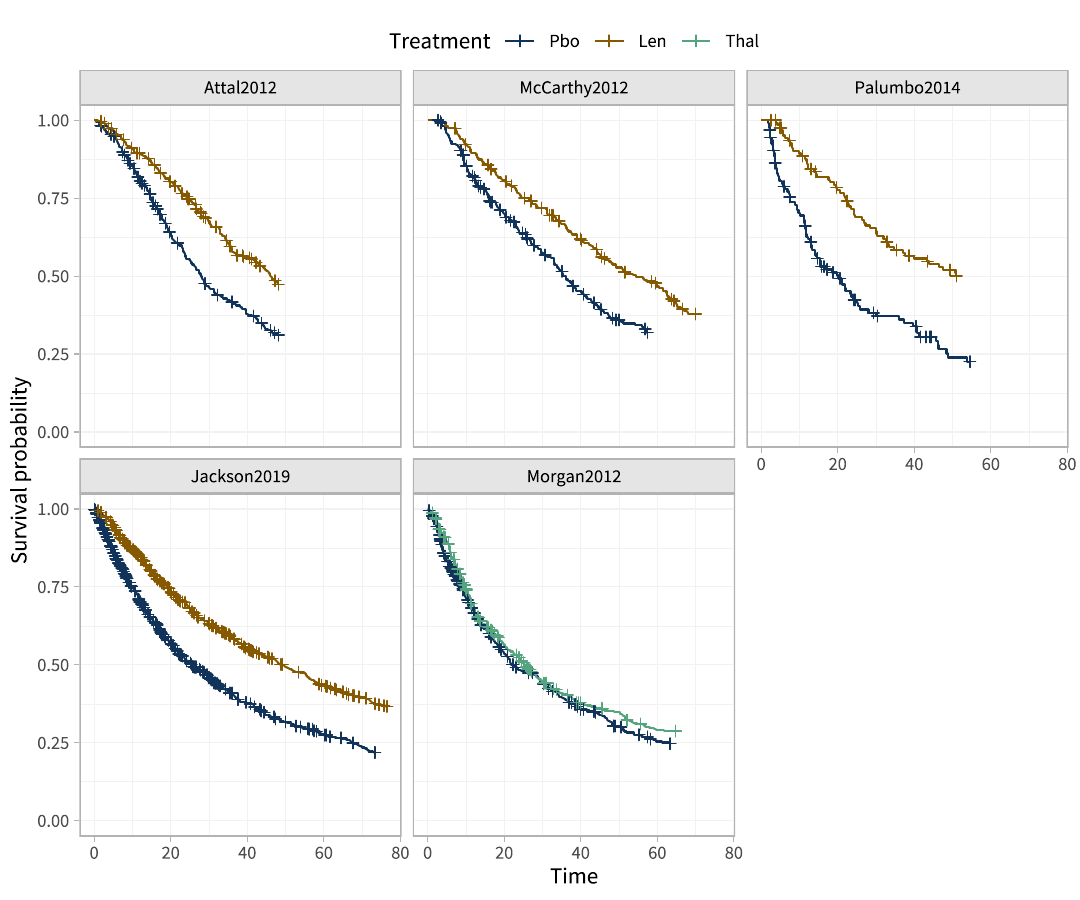}
  \caption{Kaplan-Meier curves for progression free survival on each treatment, in each study population, from the synthetic data.}
  \label{fig:ndmm_surv_sim}
\end{figure}

Three of the studies in the newly diagnosed multiple myeloma example first introduced by \textcite{Leahy2019} are analysed from available IPD.
However, since we did not have access to the original IPD from the three IPD studies, we instead constructed synthetic data that resemble the original IPD.
To do this, we first generated individual covariate values to match the published summary statistics in each study.
Binary covariates (ISS stage, response, and sex) were simulated from Binomial distributions using the reported summary proportions.
Age was simulated from Box-Cox transformed Normal distributions using the reported medians and interquartile ranges or minimum/maximum values, as implemented by the \texttt{estmeansd} R package \parencite{estmeansd}.
We then generated event times from separate generalised-F distributions in each study using the \texttt{flexsurv} R package \parencite{flexsurv}, introducing treatment and covariate effects and interactions through a model on the location parameter (log hazard rate).
The parameters of the generalised-F distributions were obtained by fitting generalised-F distributions to event/censoring times, reconstructed from digitised Kaplan-Meier curves using the algorithm of \textcite{Guyot2012}.
Values for the effect modifier interactions for the model on the location parameter were taken from those reported by \textcite{Leahy2019}.
Finally, we applied uniform censoring to the generated event times to match the observed censoring rates in each study.
The resulting Kaplan-Meier curves for this synthetic data are shown in \cref{fig:ndmm_surv_sim}, which closely resemble those reported by the original studies.

\clearpage
\setcounter{equation}{0}
\setcounter{table}{0}
\setcounter{figure}{0}
\section{Additional tables and figures for the newly diagnosed multiple myeloma example}

\begin{table}[htbp]
	\footnotesize
	\centering
	\caption{Baseline characteristics of studies included in the ML-NMR analysis of progression-free survival after ASCT for newly diagnosed multiple myeloma. Statistics are mean and standard deviation for the continuous covariate age, and percent for the categorical covariates.}
	\label{tab:ndmm_baseline_covariates}
	\begin{tabular}{lccccc}
		\toprule
		Study / Treatment & Sample size & Age (years) & ISS Stage III (\%) & Response CR/VGPR (\%) & Male (\%) \\ 
  \midrule
Attal2012 &  &  &  &  &  \\ 
  \quad Placebo & $307$ & $54.22$ $(5.24)$ & $15.96$ & $54.07$ & $57.98$ \\ 
  \quad Lenalidomide & $307$ & $54.35$ $(6.06)$ & $23.78$ & $54.72$ & $55.37$ \\ 
   [0.5ex]McCarthy2012 &  &  &  &  &  \\ 
  \quad Placebo & $229$ & $57.39$ $(5.56)$ & $18.34$ & $71.18$ & $55.46$ \\ 
  \quad Lenalidomide & $231$ & $57.93$ $(6.33)$ & $27.27$ & $62.34$ & $52.38$ \\ 
   [0.5ex]Palumbo2014 &  &  &  &  &  \\ 
  \quad Placebo & $125$ & $54.44$ $(8.98)$ & $12.00$ & $38.40$ & $63.20$ \\ 
  \quad Lenalidomide & $126$ & $53.90$ $(9.69)$ & $10.32$ & $42.06$ & $46.03$ \\ 
   [0.5ex]Jackson2019 &  &  &  &  &  \\ 
  \quad Placebo & $864$ & $64.63$ $(9.40)$ & $19.21$ & $83.10$ & $62.15$ \\ 
  \quad Lenalidomide & $1137$ & $65.17$ $(8.94)$ & $24.80$ & $82.59$ & $61.65$ \\ 
   [0.5ex]Morgan2012 &  &  &  &  &  \\ 
  \quad Placebo & $410$ & $63.92$ $(9.01)$ & $36.34$ & $71.71$ & $61.95$ \\ 
  \quad Thalidomide & $408$ & $65.59$ $(8.38)$ & $31.86$ & $74.51$ & $61.52$ \\ 
   [0.5ex]
		\bottomrule
	\end{tabular}
\end{table}

\begin{table}[htbp]
	\footnotesize
	\centering
	\caption{Estimated population-average median survival times and 95\% Credible Intervals in each study population from the cubic M-spline model.}
	\label{tab:ndmm_median_survival}
	\begin{tabular}{lccc}
		\toprule
		Study & Placebo & Lenalidomide & Thalidomide \\
		\midrule
		 Attal2012 & $28.96$ & $46.73$ & $32.30$ \\ 
   & ($26.11$, $31.94$) & ($42.09$, $51.45$) & ($26.30$, $39.63$) \\ 
   [0.5ex]McCarthy2012 & $33.84$ & $55.75$ & $38.52$ \\ 
   & ($30.02$, $37.92$) & ($49.62$, $62.34$) & ($31.16$, $46.91$) \\ 
   [0.5ex]Palumbo2014 & $22.37$ & $44.62$ & $27.59$ \\ 
   & ($18.44$, $27.02$) & ($36.24$, $54.36$) & ($20.24$, $36.90$) \\ 
   [0.5ex]Jackson2019 & $24.45$ & $50.56$ & $31.26$ \\ 
   & ($21.97$, $27.33$) & ($45.76$, $55.54$) & ($24.55$, $39.48$) \\ 
   [0.5ex]Morgan2012 & $21.25$ & $49.04$ & $27.61$ \\ 
   & ($17.80$, $25.36$) & ($38.46$, $63.79$) & ($23.13$, $32.60$) \\ 
   [0.5ex]
		\bottomrule
	\end{tabular}
\end{table}

\begin{table}[htbp]
	\footnotesize
	\centering
	\caption{Model comparison statistics for cubic M-spline models with seven and ten internal knots. The leave-one-out information criterion (LOOIC) is equal to $-2\cdot\textrm{ELPD}$, where ELPD is the expected log pointwise predictive density, and lower LOOIC values indicate better expected predictive performance. $p_\mathrm{LOO}$ is the effective number of parameters. Negative values of the ELPD difference favour the model with seven knots. Standard errors for each statistic are given alongside in small brackets.}

	\label{tab:ndmm_looic_comparison}
	\begin{tabular}{lr@{\hspace*{0.5\tabcolsep}}>{\tiny}rr@{\hspace*{0.5\tabcolsep}}>{\tiny}r}
		\toprule
		& \multicolumn{2}{c}{7 knots} & \multicolumn{2}{c}{10 knots} \\
		\midrule
		 LOOIC & $24797.7$ & ($232.2$) & $24799.2$ & ($232.1$) \\ 
  ELPD & $-12398.9$ & ($116.1$) & $-12399.6$ & ($116.1$) \\ 
  $p_\mathrm{LOO}$ & $35.1$ & ($0.7$) & $37.4$ & ($0.7$) \\ 
  ELPD difference &  &  & $-0.8$ & ($1.1$) \\ 
  
		\bottomrule
	\end{tabular}
\end{table}

\begin{table}[htbp]
	\footnotesize
	\centering
	\caption{LOOIC model comparison statistics for cubic M-spline models with seven and ten internal knots, computed within each study in the network.}

	\label{tab:ndmm_looic_by_study}
	\begin{tabular}{lrr}
		\toprule
		Study & 7 knots & 10 knots \\ 
  \midrule
Attal2012 & $3345.1$ & $3346.6$ \\ 
  Jackson2019 & $12398.7$ & $12400.3$ \\ 
  McCarthy2012 & $2726.4$ & $2724.5$ \\ 
  Morgan2012 & $4990.7$ & $4991.3$ \\ 
  Palumbo2014 & $1336.8$ & $1336.6$ \\ 
  
		\bottomrule
	\end{tabular}
\end{table}

\begin{figure}[htbp]
	\centering
	\includegraphics[width=\textwidth]{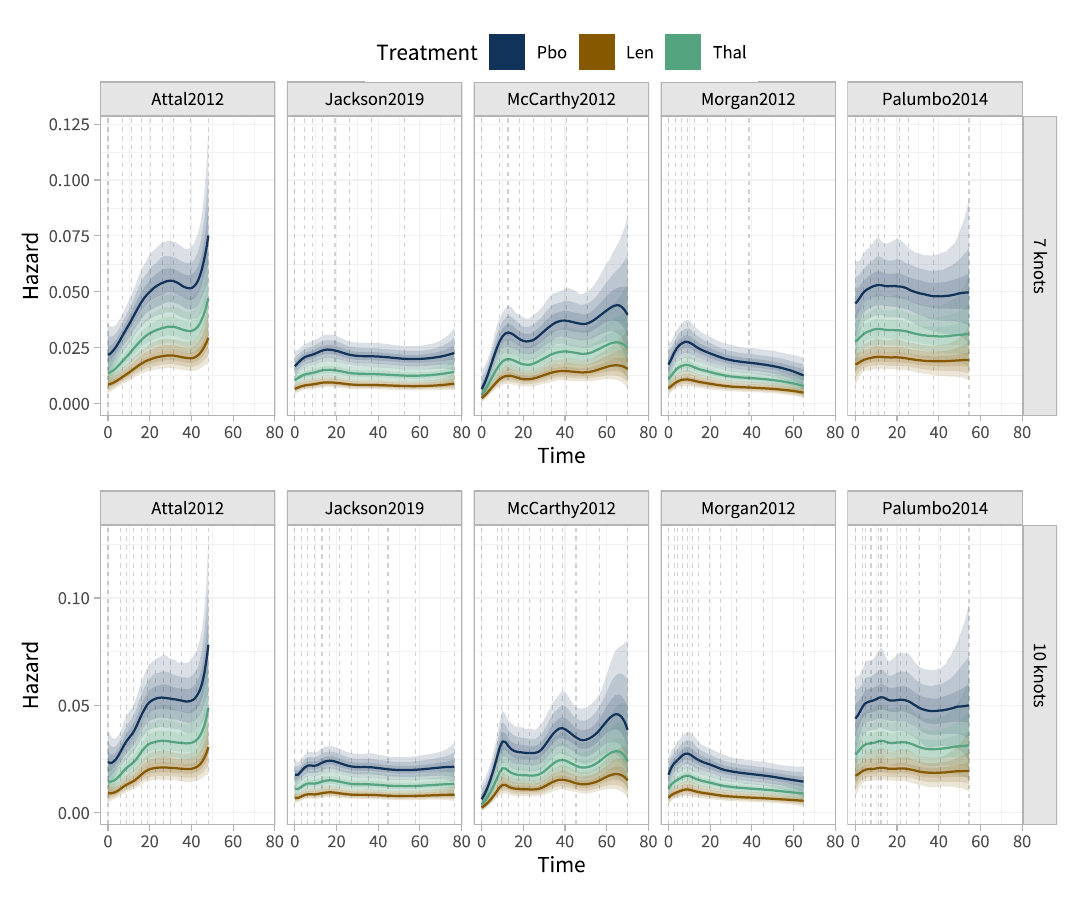}
	\caption{Estimated baseline hazard curves on each treatment in each study population, under cubic M-spline models with seven and ten internal knots. These are individual-level conditional hazards at the reference level of the covariates: female, ISS stage I-II, no complete/very good partial response, age 61.7.  Shaded bands indicate the 50\%, 80\%, and 95\% Credible Intervals for the hazard curves (thick lines). Knot locations are indicated by the vertical dashed lines.}
	\label{fig:ndmm_baseline_hazard_est_curves}
\end{figure}

\begin{figure}[htbp]
	\centering
	\includegraphics[width=\textwidth]{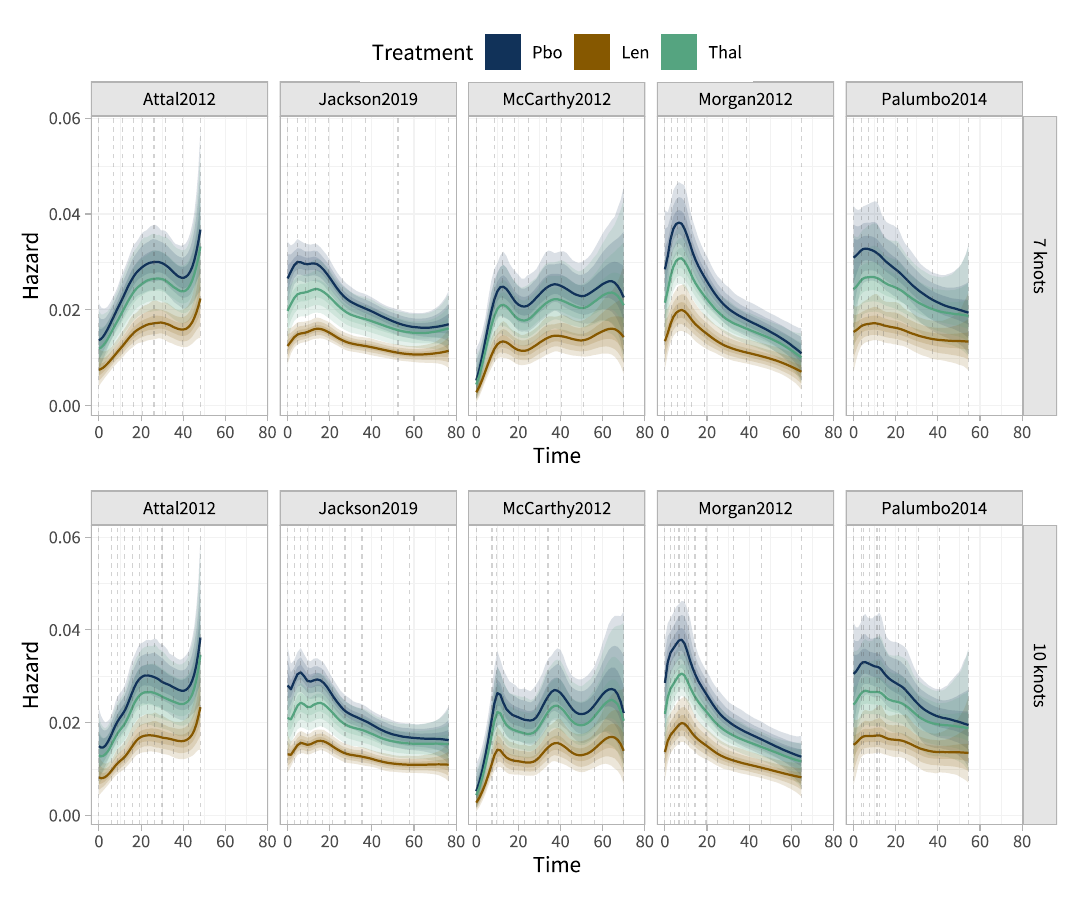}
	\caption{Estimated marginal hazard curves on each treatment in each study population, under cubic M-spline models with seven and ten internal knots. Shaded bands indicate the 50\%, 80\%, and 95\% Credible Intervals for the hazard curves (thick lines). Knot locations are indicated by the vertical dashed lines.}
	\label{fig:ndmm_hazard_est_curves}
\end{figure}

\clearpage

\begin{table}[htbp]
	\footnotesize
	\centering
	\caption{Model comparison statistics for ML-NMR models with and without the proportional hazards assumption. The leave-one-out information criterion (LOOIC) is equal to $-2\cdot\textrm{ELPD}$, where ELPD is the expected log pointwise predictive density, and lower LOOIC values indicate better expected predictive performance. $p_\mathrm{LOO}$ is the effective number of parameters. Negative values of the ELPD difference favour the proportional hazards model. Standard errors for each statistic are given alongside in small brackets.}

	\label{tab:ndmm_looic_nph}
	\begin{tabular}{lr@{\hspace*{0.5\tabcolsep}}>{\tiny}rr@{\hspace*{0.5\tabcolsep}}>{\tiny}r}
		\toprule
		& \multicolumn{2}{c}{PH} & \multicolumn{2}{c}{Non-PH} \\
		\midrule
		 LOOIC & $24797.7$ & ($232.2$) & $24811.0$ & ($232.3$) \\ 
  ELPD & $-12398.9$ & ($116.1$) & $-12405.5$ & ($116.1$) \\ 
  $p_\mathrm{LOO}$ & $35.1$ & ($0.7$) & $44.1$ & ($0.8$) \\ 
  ELPD difference &  &  & $-6.6$ & ($3.5$) \\ 
  
		\bottomrule
	\end{tabular}
\end{table}

\begin{table}[htbp]
	\footnotesize
	\centering
	\caption{LOOIC model comparison statistics for ML-NMR models with and without the proportional hazards assumption, computed within each study in the network.}

	\label{tab:ndmm_looic_nph_by_study}
	\begin{tabular}{lrr}
		\toprule
		Study & PH & Non-PH \\ 
  \midrule
Attal2012 & $3345.1$ & $3347.2$ \\ 
  Jackson2019 & $12398.7$ & $12399.9$ \\ 
  McCarthy2012 & $2726.4$ & $2738.0$ \\ 
  Morgan2012 & $4990.7$ & $4989.6$ \\ 
  Palumbo2014 & $1336.8$ & $1336.3$ \\ 
  
		\bottomrule
	\end{tabular}
\end{table}

\begin{table}[htbp]
	\footnotesize
	\centering
	\caption{Model comparison statistics for unadjusted NMA models with and without the proportional hazards assumption. The leave-one-out information criterion (LOOIC) is equal to $-2\cdot\textrm{ELPD}$, where ELPD is the expected log pointwise predictive density, and lower LOOIC values indicate better expected predictive performance. $p_\mathrm{LOO}$ is the effective number of parameters. Negative values of the ELPD difference favour the proportional hazards model. Standard errors for each statistic are given alongside in small brackets.}

	\label{tab:ndmm_looic_nma}
	\begin{tabular}{lr@{\hspace*{0.5\tabcolsep}}>{\tiny}rr@{\hspace*{0.5\tabcolsep}}>{\tiny}r}
		\toprule
		& \multicolumn{2}{c}{PH} & \multicolumn{2}{c}{Non-PH} \\
		\midrule
		 LOOIC & $24947.1$ & ($230.5$) & $24951.9$ & ($230.6$) \\ 
  ELPD & $-12473.5$ & ($115.2$) & $-12475.9$ & ($115.3$) \\ 
  $p_\mathrm{LOO}$ & $27.2$ & ($0.4$) & $37.6$ & ($0.6$) \\ 
  ELPD difference &  &  & $-2.4$ & ($4.4$) \\ 
  
		\bottomrule
	\end{tabular}
\end{table}

\begin{table}[htbp]
	\footnotesize
	\centering
	\caption{LOOIC model comparison statistics for unadjusted NMA models with and without the proportional hazards assumption, computed within each study in the network.}

	\label{tab:ndmm_looic_nma_by_study}
	\begin{tabular}{lrr}
		\toprule
		Study & PH & Non-PH \\ 
  \midrule
Attal2012 & $3410.3$ & $3410.5$ \\ 
  Jackson2019 & $12404.7$ & $12398.7$ \\ 
  McCarthy2012 & $2770.3$ & $2781.4$ \\ 
  Morgan2012 & $4989.1$ & $4990.6$ \\ 
  Palumbo2014 & $1372.7$ & $1370.6$ \\ 
  
		\bottomrule
	\end{tabular}
\end{table}

\end{document}